\documentclass[useAMS,usenatbib]{mn2e}
\usepackage{epsfig}
\usepackage{graphicx}
\usepackage{amssymb}
\usepackage{color}

\title[An oblique pulsar magnetosphere with a plasma conductivity]{An oblique pulsar magnetosphere with a plasma conductivity}
\author[Cao et al. ]
{Gang Cao, Li Zhang \thanks{E-mail: lizhang@ynu.edu.cn} and
Sineng Sun
\\
Department of Astronomy, Yunnan University, Key Laboratory of Astroparticle Physics of Yunnan Province, Kunming, 650091, China\\
}

\begin{document}
\pagerange{\pageref{firstpage}--\pageref{lastpage}} \pubyear{2016}

\maketitle

\label{firstpage}

\begin{abstract}
An oblique pulsar magnetosphere with a plasma conductivity is studied by using a pseudo-spectral method. In the pseudo-spectral method, the time-dependent Maxwell equations are solved, both electric and magnetic fields are expanded in terms of the vector spherical harmonic (VSH) functions in spherical geometry and the divergencelessness of magnetic field is analytically enforced by a projection method. The pulsar magnetospheres in infinite (i. e., force-free approximation) and finite conductivities are simulated and  a family of solutions that smoothly transition from the Deutsch vacuum solution to the force-free solution are obtained. The $\sin^2 \alpha$ dependence of the spin-down luminosity on the magnetic inclination angle $\alpha$ in which the full electric current density are taken into account is retrieved
in the force-free regime.
\end{abstract}


\section{Introduction}

Pulsars are rotating neutron stars with very strong magnetic fields of order $\sim 10^{12}$ G. They can produce the electromagnetic radiation throughout almost the entire electromagnetic spectrum from radio to $\gamma$-ray bands. The radiations from these objects originate from high-energy charged particles accelerated  along the open magnetic field line by the electric field induced by the rotating magnetic field.
The emission from pulsars provides the valuable information about pulsar magnetospheric structure, particle acceleration mechanism, and plasma physics in strong magnetic fields.

The structure of realistic pulsar magnetosphere is still an unresolved problem. In the early pulsar research, the pulsar magnetosphere is generally assumed to be a static (or retarded) vacuum dipole magnetic field in various radiation models of pulsars, which is described by a analytical formula given by \citet{deu55}. Both outer gap \citep[e.g.,][]{che86,zha97,che00} and slot gap \citep[e.g.,][]{mus04,har08} models are based on this field structure, and such models has been met with great success in modeling pulsar GeV emission and light curves \citep[e.g.,][]{zha02,lz10,ljz13,jia14}. However, the vacuum solution has no plasma, it is impossible to produce any pulsar emission. In fact, pulsar magnetosphere is known to be filled with abundant plasma \citep{gol69}. This plasma shorts out the accelerating electric field to form a force-free magnetosphere.
Follow the work of \citet{gol69}, many attempts are made to produce the structure of the pulsar magnetosphere.  Even in the simplest axisymmetric case, the pulsar magnetosphere still remains uncertain for a long time. An significant advance was made by \citet{con99}. The CKF solution consists of a region of closed field lines extending to the light cylinder, an open zone with asymptotically monopole poloidal field lines and an equatorial current sheet beyond the light cylinder.
It was later found that the steady-state solutions with the Y-points exist within the light cylinder \citep{goo04,tim06}. This raises the question of the uniqueness of the CKF solution. The best way of solving this problem is to solve the time-dependent Maxwell equations. The time-dependent simulations for the force-free aligned rotator  have been studied by several authors based on different methods \citep[e.g.,][]{kom06,mck06,tim06,yu11,par12,cao16}. The axisymmetric solutions generally converge to a similar CKF solution and reveal an equatorial current sheet beyond the light cylinder.

3D simulations of pulsar magnetosphere have achieved significant progress in recent years. \citet{spi06} used the finite-difference time-domain method  to solve the time-dependent Maxwell equations in the Cartesian coordinate and presented the first 3D structure of pulsar magnetosphere for the oblique rotator. \citet{kal09} improved the method of \citet{spi06} by implementing a perfectly matched layer at the outer boundaries, which allow the system to evolve for many stellar periods to reach a final steady state.
However, it is difficult to impose the exact inner boundary at the stellar surface in the Cartesian coordinate.
\citet{pet12} developed a pseudo-spectral method in the spherical coordinate to improve this situation and confirmed the structure of 3D force-free solutions.
Meanwhile, these studies have been extended to the full magnetohydrodynamic (MHD) regime that takes plasma pressures and inertial into account \citep{tch13}, and the general-relativistic force-free regime that takes space-time curvature and frame-dragging effects into accounts \citep{pet16}.
It should be noted that the vacuum solution has accelerating electric fields but has no particles , while the force-free solution has enough particles to shout out the accelerated electric fields. Therefore, they cannot provide any information about the sites of particle acceleration and radiation. The realistic pulsar magnetosphere
should therefore lie between the vacuum limit and force-free limit. The force-free approximation cannot allow any accelerating electric field. A way to introduce accelerating electric fields in the magnetosphere is to allow finite resistivity of the plasma. Two groups constructed a sets of resistive solutions that smoothly bridges the gap between the vacuum and force-free solutions \citep{li12a,kal12}.
The resistive solutions have important implication on the intermittent pulsar behavior and have been proposed to explain explain the ``on/off'' state of the intermittent pulsar \citep{li12b,kal12,con14}. Moreover,
the resistive solutions have been used to model the Fermi $\gamma$-ray spectra and light curves \citep{kal14,bra15}.
Recently, particle-in-cell (PIC) methods are also used to simulate the structure of pulsar magnetosphere
that takes the self-consistent dynamics of particles and fields into accounts \citep{phi14,che14,bel15,cer15,phi15}.

In the previous paper, we presented the time-dependent simulation of the force-free axisymmetric pulsar magnetosphere by the pseudo-spectral method \citep{cao16}. Here, we extend our pseudo-spectral code to 3D geometry by including the Fourier transform in the azimuthal direction. Then, we present the  3D simulations of the oblique rotator with a plasma conductivity $\sigma$, where the Deutsch vacuum solution ($\sigma=0$) and the  force-free solution ($\sigma \rightarrow \infty$) are considered as two extreme cases.
The outline of this paper is as follows. In \S 2 we describe the force-free and resistive electrodynamics, and in \S 3 we briefly describe our numerical algorithm. We present the structure of pulsar magnetosphere for an oblique rotator with the force-free and resistive electrodynamics in \S 4. Finally, we give our discussion and conclusion in \S 5.

\section{FORCE-FREE and RESISTIVE ELECTRODYNAMICS}

The time-dependent Maxwell equations are
\begin{eqnarray}
{\partial {\bf B}\over \partial t}&=&-{\bf \nabla} \times {\bf E}\;,\\
{\partial  {\bf E}\over \partial t}&=&{\bf \nabla} \times {\bf B}-{\bf J}\;,
\label{Eq1-2}
\end{eqnarray}
supplemented with two initial condition
\begin{eqnarray}
\nabla\cdot{\bf B}&=&0\;, \\
\nabla\cdot{\bf E}&=&\rho_{\rm e}\;,
\label{Eq3-4}
\end{eqnarray}
where $\rho_{\rm e}=\nabla\cdot{\bf E}$ is the charge density and ${\bf J}$ is the current density. Note that we have set the light speed $c=1$ and $4\pi=1$ throughout this paper.
The force-free approximation implies negligible partial inertial and pressure. Therefore, the Lorentz force acting on a plasma fluid element must
vanish,
\begin{eqnarray}
\rho_{\rm e}{\bf E}+\bf J\times\bf B=0\;,
\label{Eq5}
\end{eqnarray}
which implies the force-free condition $\bf E\cdot \bf B=0$.
From the force-free condition and the Maxwell equations, the current sheet is uniquely determined to be \citep{gru99,bla02}
\begin{eqnarray}
{\bf J}= \rho_{\rm e} {{\bf E} \times {\bf B} \over B^2}+
{({\bf B}\cdot {\bf \nabla}\times {\bf B}-{\bf E}\cdot {\bf \nabla}\times {\bf E}){\bf{B}}
\over B^2}\;.
\label{Eq6}
\end{eqnarray}
The first term is the drift current perpendicular to $\bf B$, and the second term is conduction current parallel to  $\bf B$, which maintains the force-free condition.

Note that the full electric current expression is not taken into account in the finite-difference schemes. The field-aligned component is usually dropped because of the intricate expression including spatial derivatives. It is tricky to handle with finite-difference schemes, which needs the interpolation of both field and field derivatives. However, it is easy for the spectral method to handle with spatial derivatives. Therefore, we include the full electric current expression in our pseudo-spectral code.

The force-free solutions do not allow for any electric field parallel to ${\bf {B}}$ (${\bf {E}}_{\|}=0$). They do not provide any possibility of particle
acceleration and production of radiation. Therefore, a new degree of freedom is needed to describe the realistic pulsar magnetosphere. The resistivity of plasma can allow for ${\bf {E}}_{\|} \neq 0 $, which is required for the particle acceleration and produces the observed radiation.
We use the following formula to describe the current density  \citep{kal14}
\begin{eqnarray}
{\bf J}= \rho_{\rm e} {{\bf E} \times {\bf B} \over B^2+E^2_{0}}+\sigma {\bf {E}}_{\|}\;,
\label{Eq6}
\end{eqnarray}
where $\sigma$ is plasma conductivity measured in unit of the pulsar rotation frequency $\Omega$, $E_{0}$ is defined by $B^2_{0}-E^2_{0}={\bf B}^2-{\bf E}^2$, $E_{0}B_{0}=\bf E\cdot \bf B$, $E_{0}\geq0$, and ${\bf {E}}_{\|}=({\bf E}\cdot {\bf B}) {\bf B}/B^2$ is the electric field parallel to the magnetic field.
The first term in equation (7) is the drift component, while the second term controls ${\bf {E}}_{\|}$ by the conductivity $\sigma$.
The $E_{0}$ term in equation (7) prevents the magnetic field going to zero in the current sheet and thus ensure the drift current to be superluminal.
As $\sigma$ ranges from $\sigma=0$ to $\sigma \rightarrow \infty$, we should obtain a family of solution from the Deutsch vacuum solution to the  force-free solution.

\section{PSEUDO-SPECTRAL ALGORITHM}

In this section, we  briefly describe the pseudo-spectral algorithm developed by \cite{pet12} and \citet{cao16}.
The main ingredients are the VSH expansion of the electric and magnetic fields, maintaining the divergencelessness constraints on {\bf B}, an exact enforcement of boundary condition, an explicit time integration with the third-order Adam-Bashforth scheme, and a spectral filtering.

In the pseudo-spectral algorithm, the electric and magnetic fields are expanded in terms of the VSHs by the expressions \citep[e.g.,][]{pet12}
\begin{eqnarray}
  \label{eq:B_VSH}
  \mathbf{B} & = & \sum_{l=0}^\infty\sum_{m=-l}^l
  \left(B^r_{lm} \mathbf{Y}_{lm} + B^{(1)}_{lm} \mathbf{\Psi}_{lm}+
    B^{(2)}_{lm}\mathbf{\Phi}_{lm}\right)\;,\\
  \label{eq:E_VSH}
  \mathbf{E} & = & \sum_{l=0}^\infty\sum_{m=-l}^l
  \left(E^r_{lm} \mathbf{Y}_{lm} + E^{(1)}_{lm}\mathbf{\Psi}_{lm}+
    E^{(2)}_{lm}\mathbf{\Phi}_{lm}\right)\;.
\end{eqnarray}
From the expansion coefficients, we can easily compute the linear differential operators like $\nabla\cdot$ and $\nabla\times$ in the coefficient space and then transformed back to the real space to advance the solution in time. For a detail discussion about the definitions and  properties of VSHs, see \cite{pet12}. In the radial coordinate, the spectral coefficients are expanded into the Chebyshev function such that
\begin{eqnarray}
B^r_{lm}(t,r)=\sum_{k=0}^{N_{r}-1}B^r_{klm}(t) T_{k}(r)\;.
\end{eqnarray}
The radial derivative can be easily computed by the three-term recurrence relation, see \citet{can06}.

In the force-free electrodynamics, the force-free evolution may develop some regions in which the force-free condition is violated. Therefore, we enforce the $\bf E \cdot \bf B=0$ and $ E<B$ conditions regularly. At each time step, we subtract the parallel electric field by adjusting the electric field as ${\bf {E}}= {\bf {E}}-{{ {\bf {E}} \cdot \bf {B} } \over {B^2}} \bf {B}$. If this new electric field does not satisfy the condition $ E<B$, we reset the electric field as $E=B$, such that ${\bf {E}} = {\bf {E}} \sqrt{{B^2}\over{E^2}}$.
In the resistive electrodynamics, the current density in equation (7) prevents magnetic field becoming zero in the current sheet. Therefore, no additional constraints are required during the time evolution. We let the system evolve following the Maxwell equations.
In the Maxwell equations, the magnetic field is divergencelessness ($\nabla \cdot {\bf B}=0$). This property can be analytically enforced by the expansion below:
\begin{equation}
{\bf {B}} = \sum_{l=1}^\infty\sum_{m=-l}^l
 \left(\nabla \times [f_{lm}^{B}(r,t) \, { {\bf{\Phi}}_{lm} }] + g_{lm}^{B}(r,t) \, {\bf{\Phi}}_{lm} \right),
\end{equation}
where $\{f^B_{lm}(r,t),g^B_{lm}(r,t)\}$ is the expansion coefficients of $\bf {B}$. This expansion is divergencelessness by definition. Therefore, to impose the divergencelessness constrain, the magnetic field is projected onto the subspace defined by equation (11) at each time step. Specifically, we perform a forward transform from the vector $\bf B$ to the coefficients $\{f^B_{lm}(r,t),g^B_{lm}(r,t)\}$  and a backward transform from the coefficients to the vector $\bf B$.

\begin{figure*}
\begin{tabular}{cccccc}
  \includegraphics[width=5.5cm,height=5.5cm]{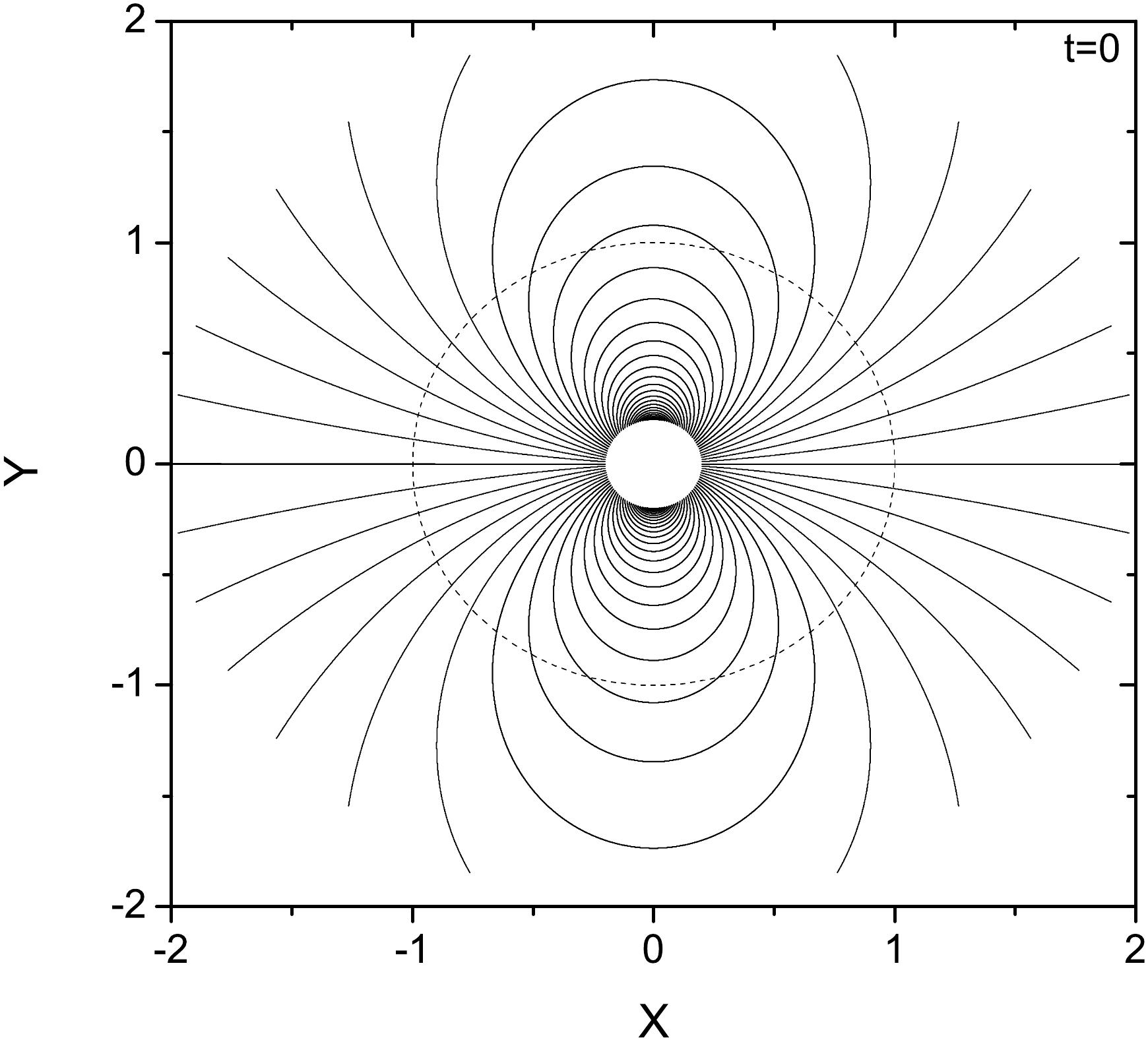} \qquad
  \includegraphics[width=5.5cm,height=5.5cm]{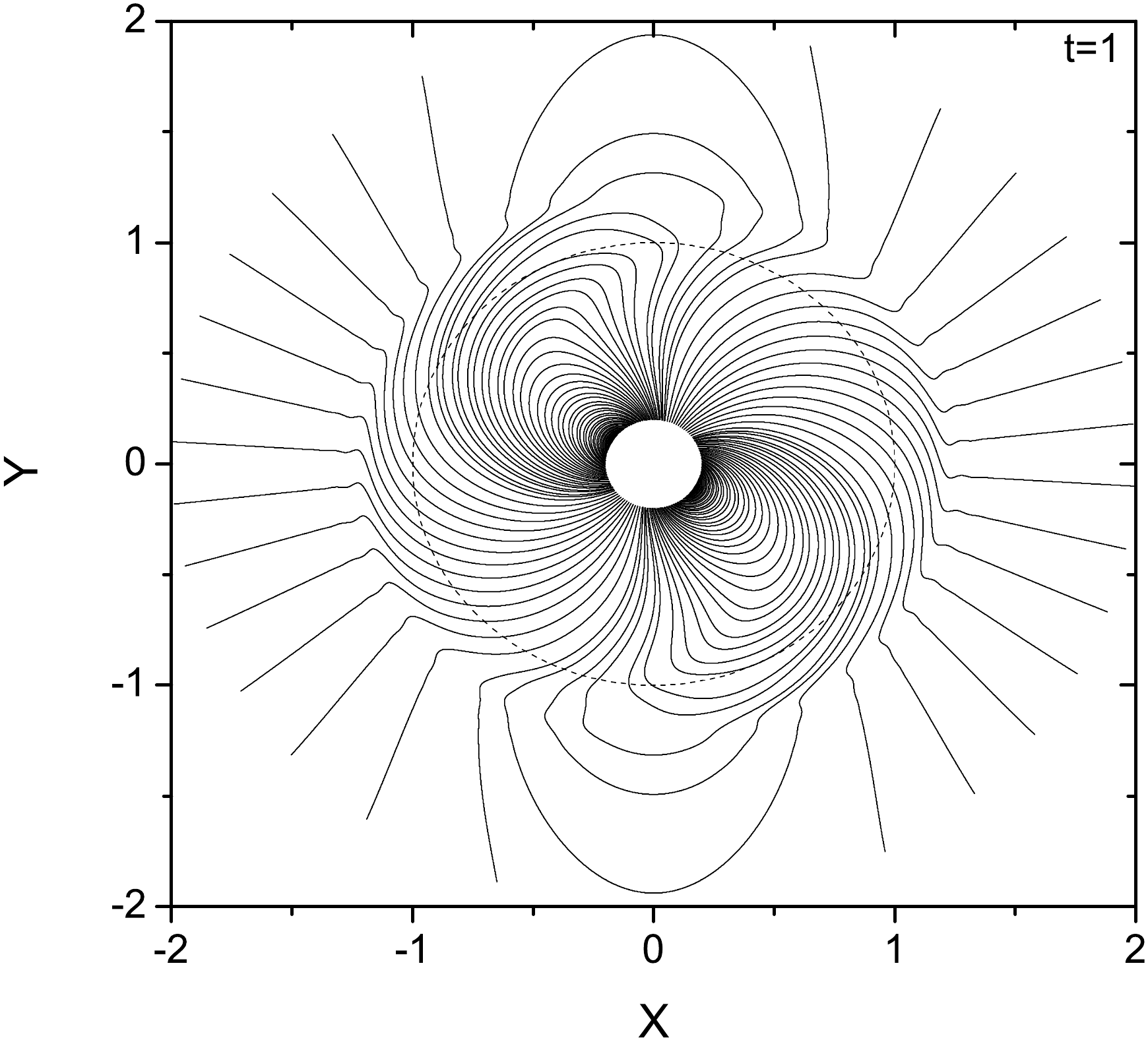} \qquad
  \includegraphics[width=5.5cm,height=5.5cm]{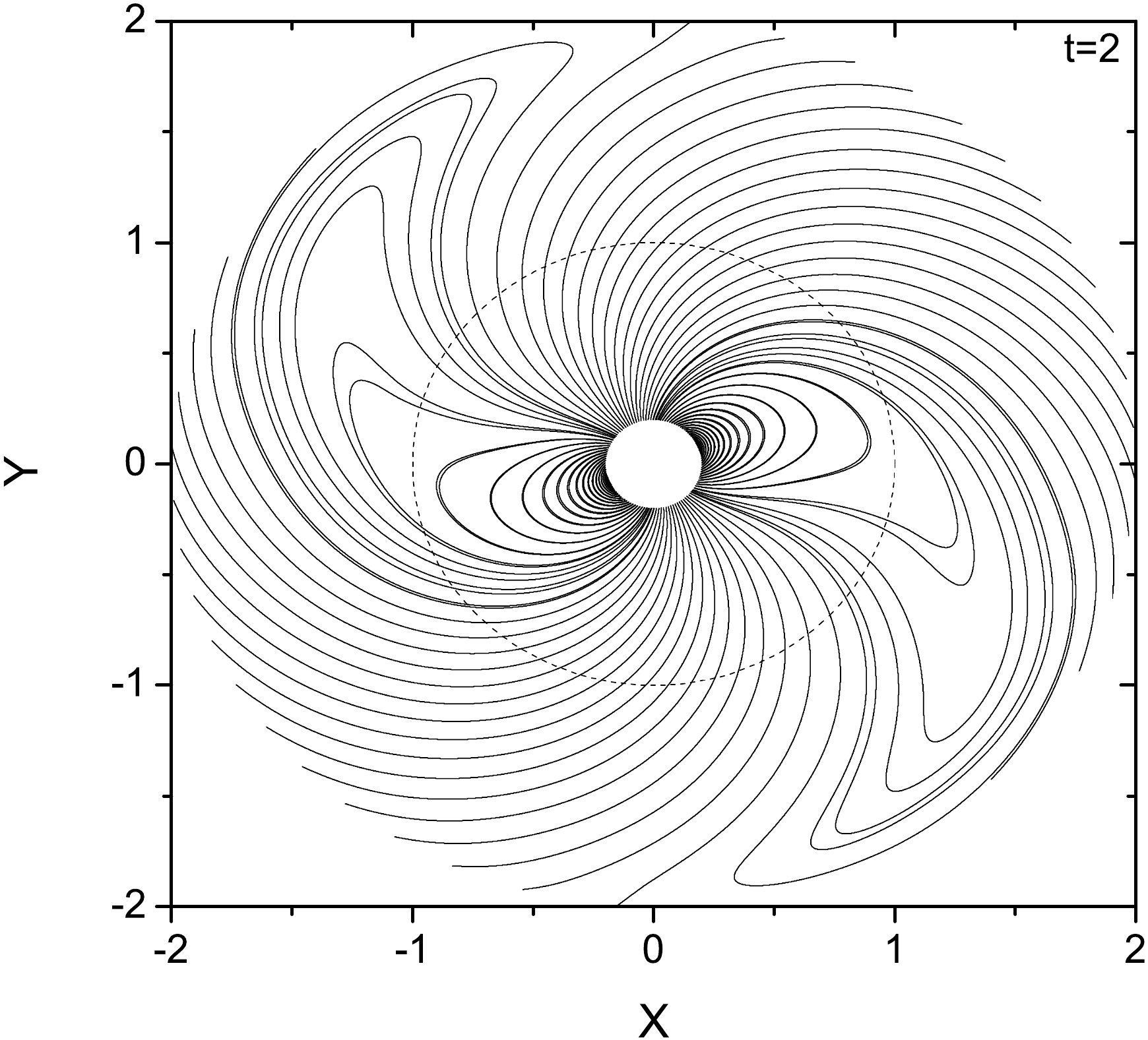} \\
  \includegraphics[width=5.5cm,height=5.5cm]{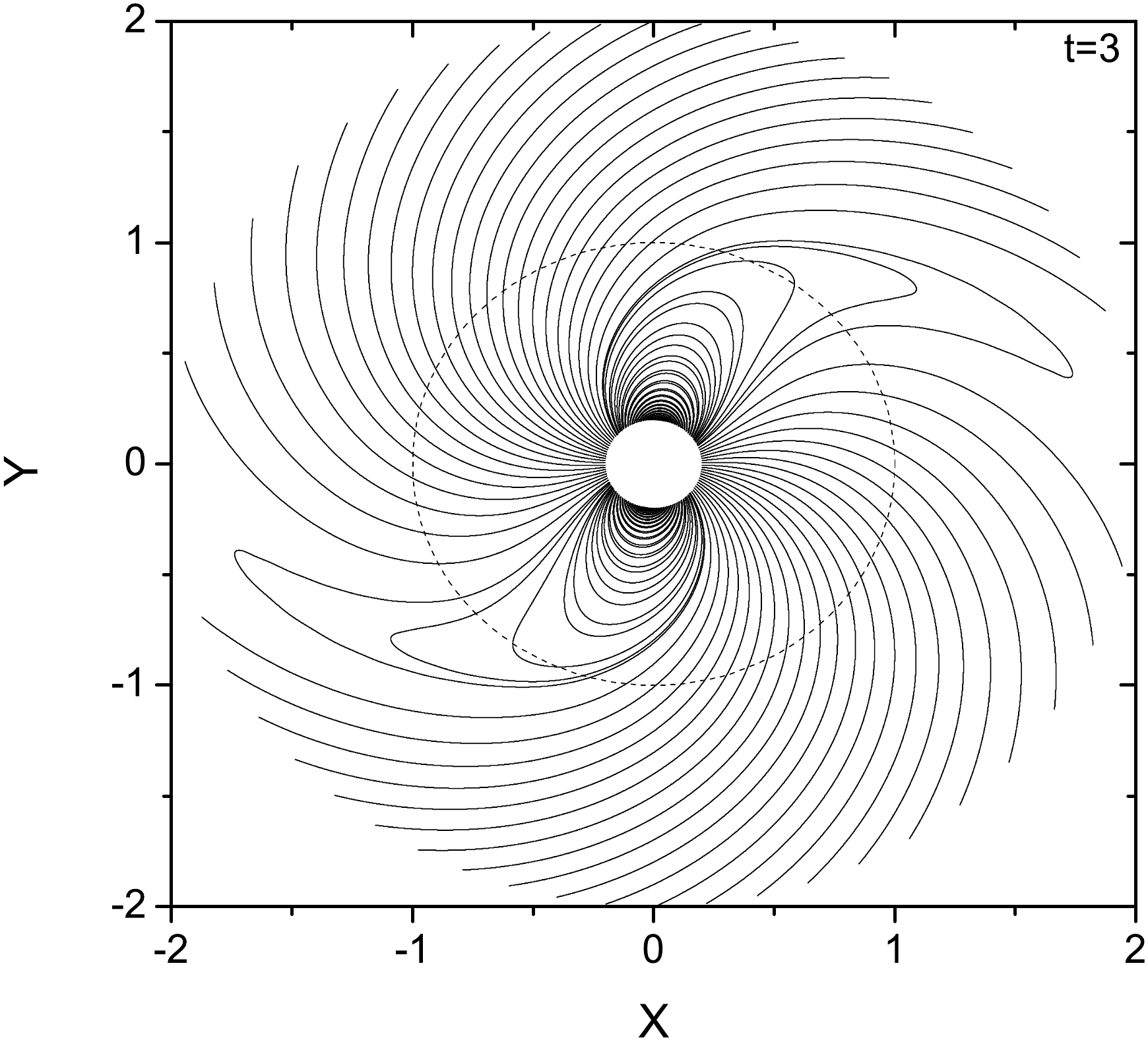} \qquad
  \includegraphics[width=5.5cm,height=5.5cm]{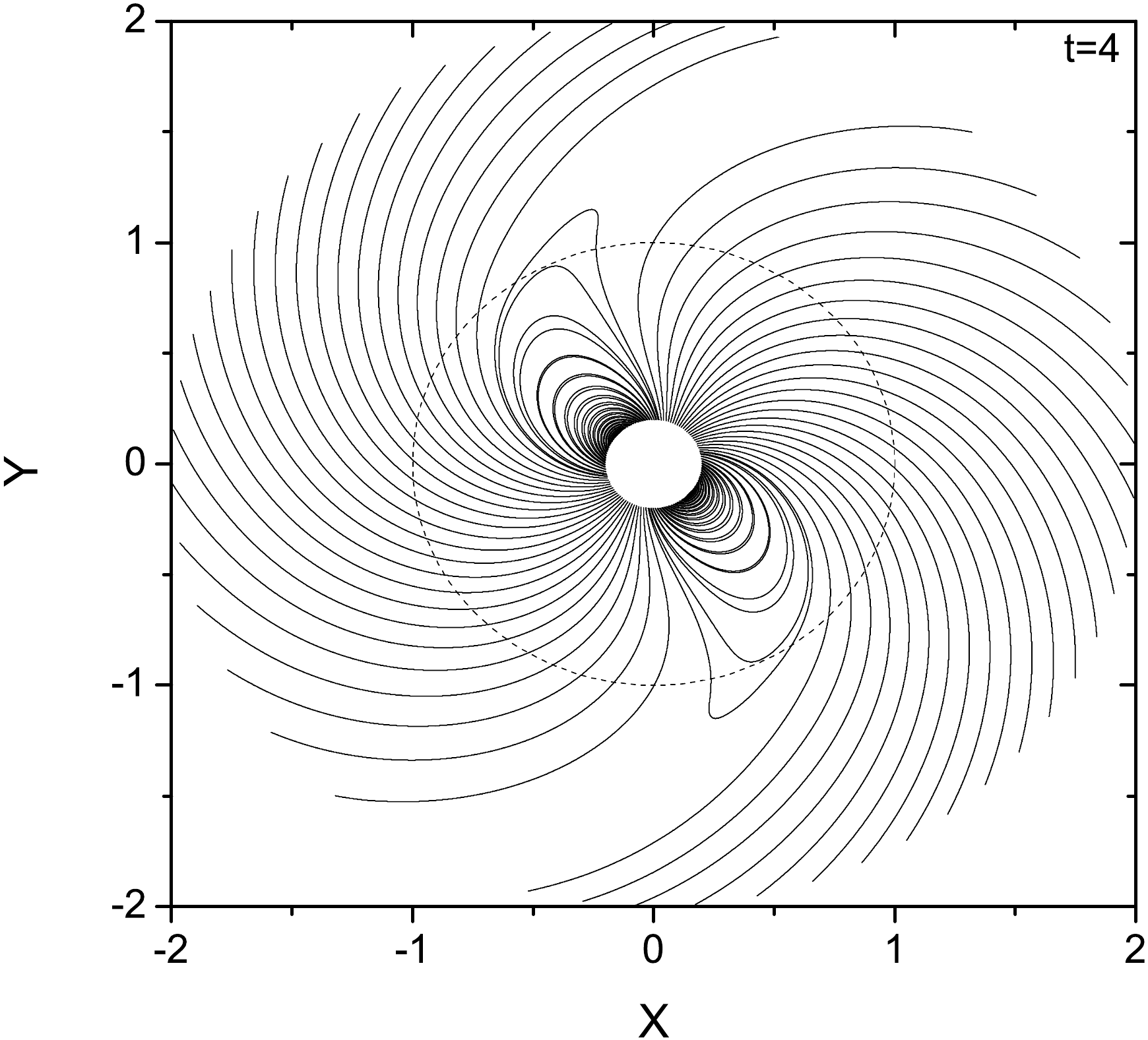} \qquad
  \includegraphics[width=5.5cm,height=5.5cm]{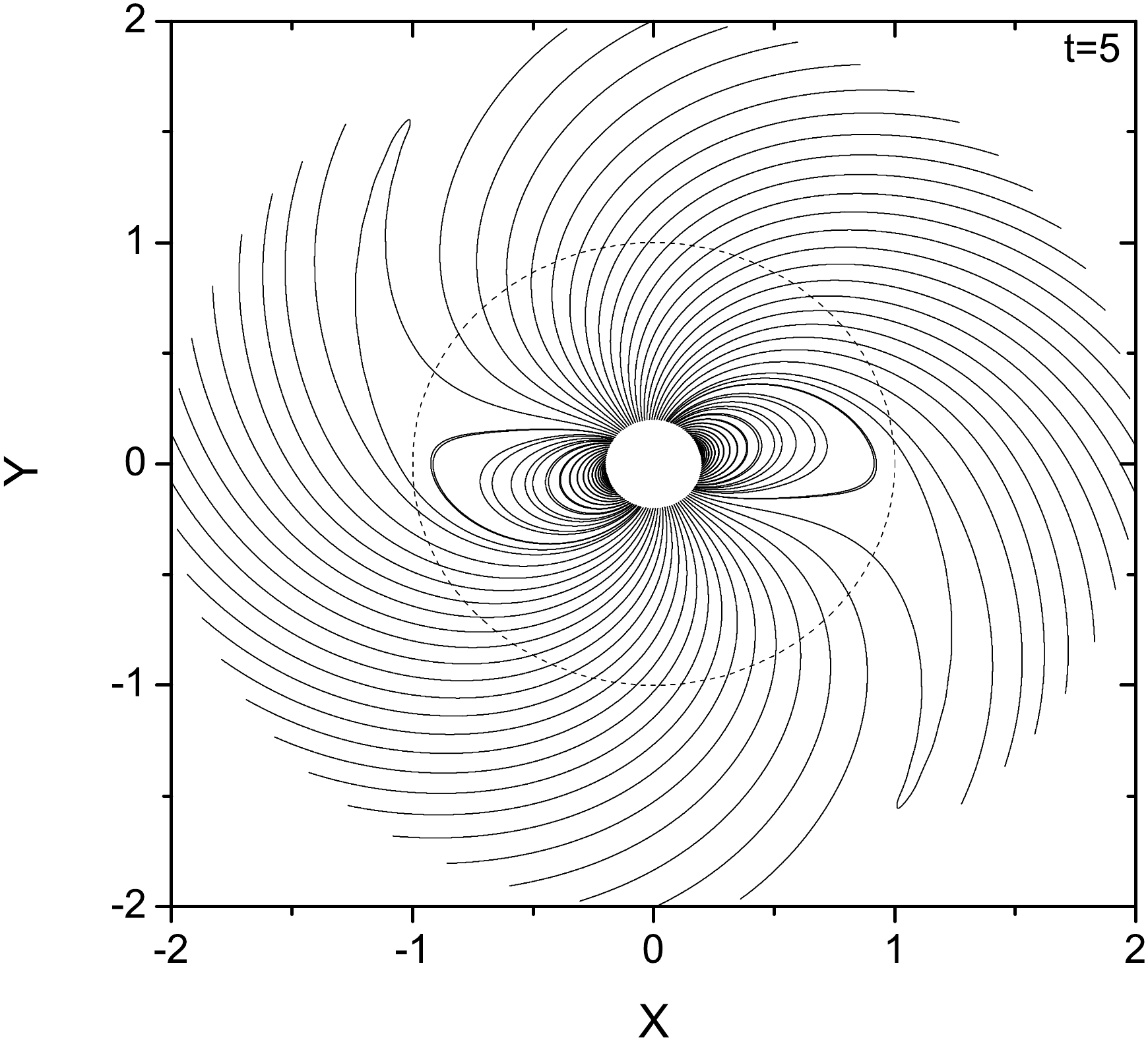} \\
  \includegraphics[width=5.5cm,height=5.5cm]{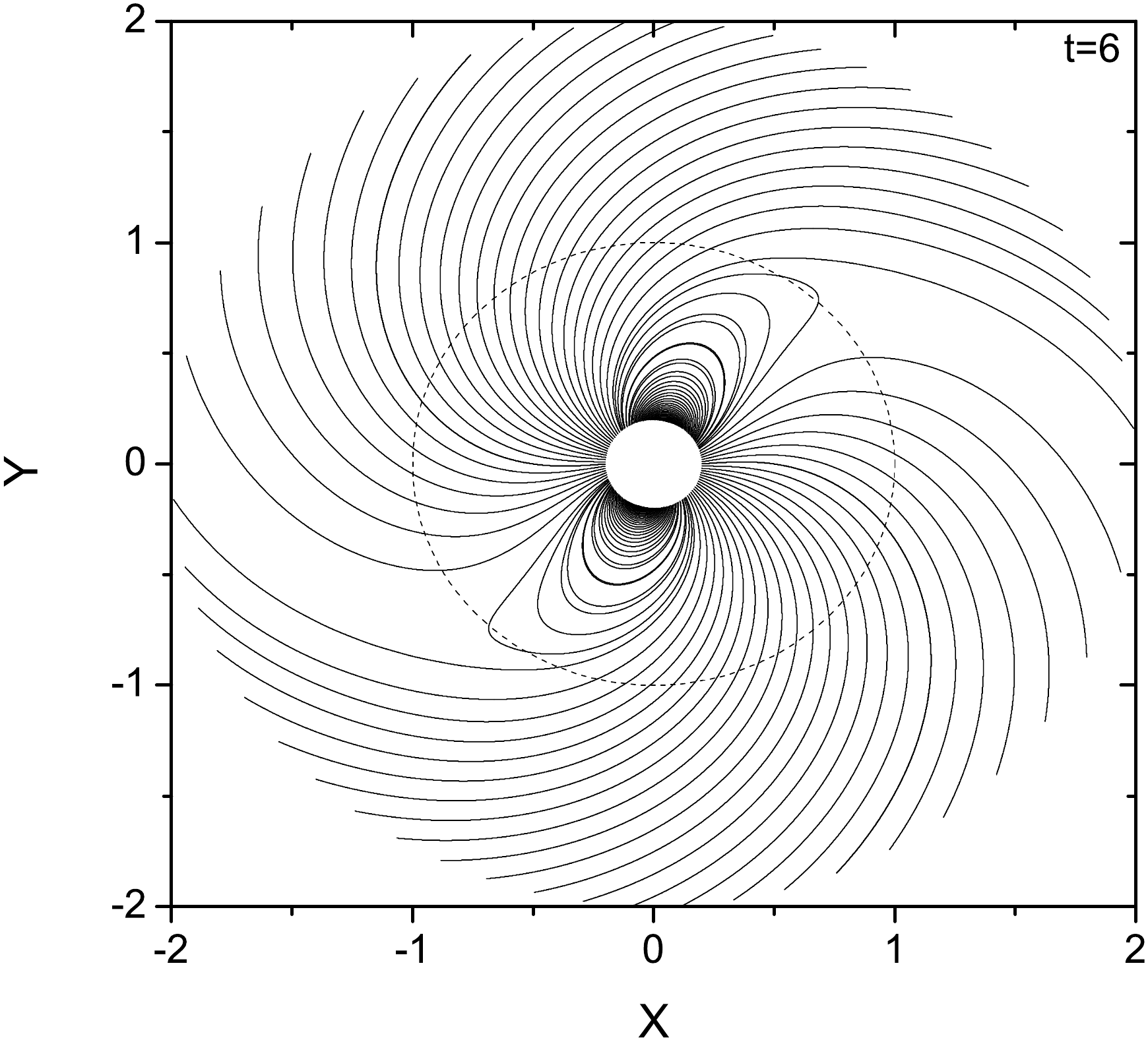} \qquad
  \includegraphics[width=5.5cm,height=5.5cm]{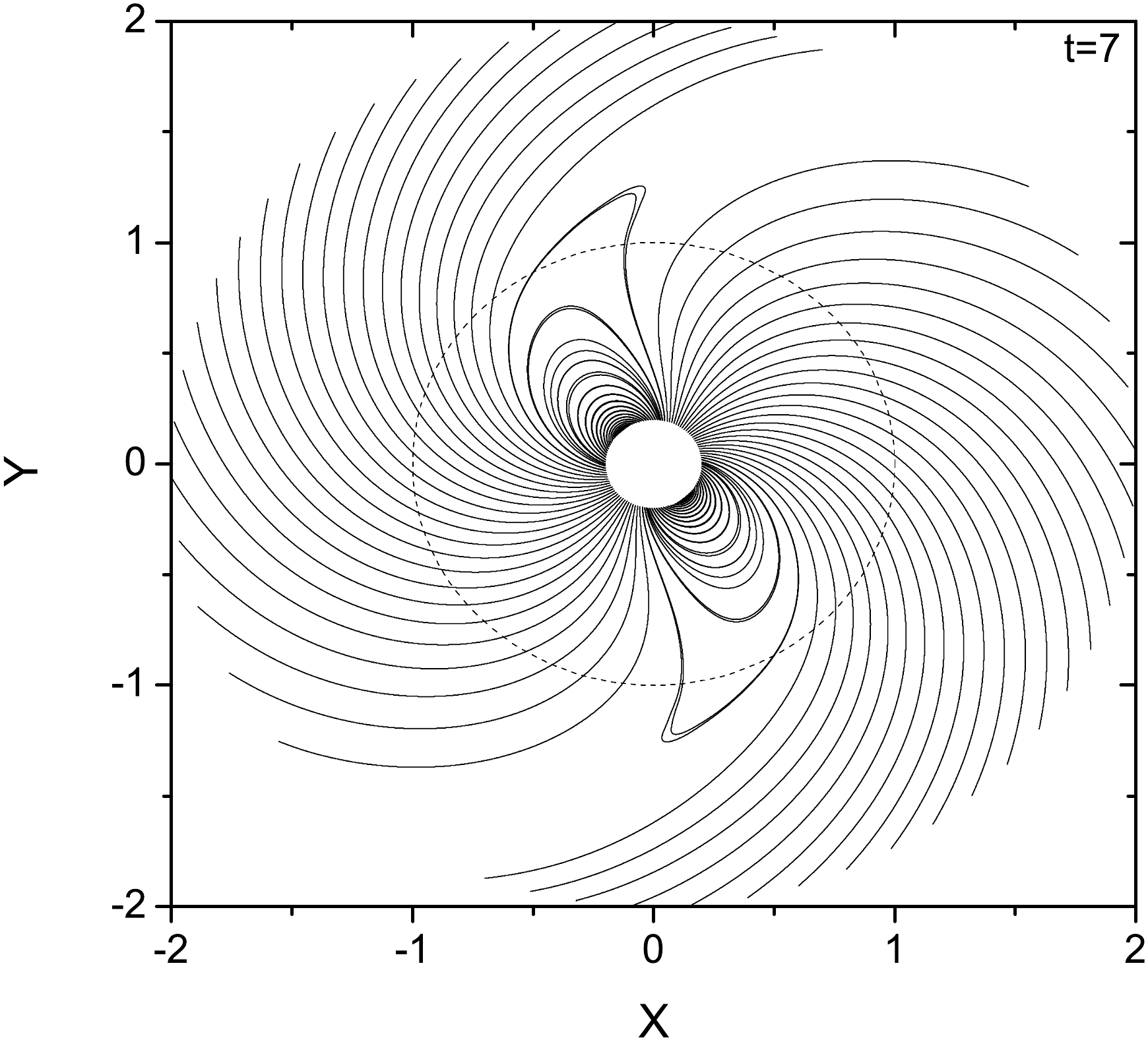} \qquad
  \includegraphics[width=5.5cm,height=5.5cm]{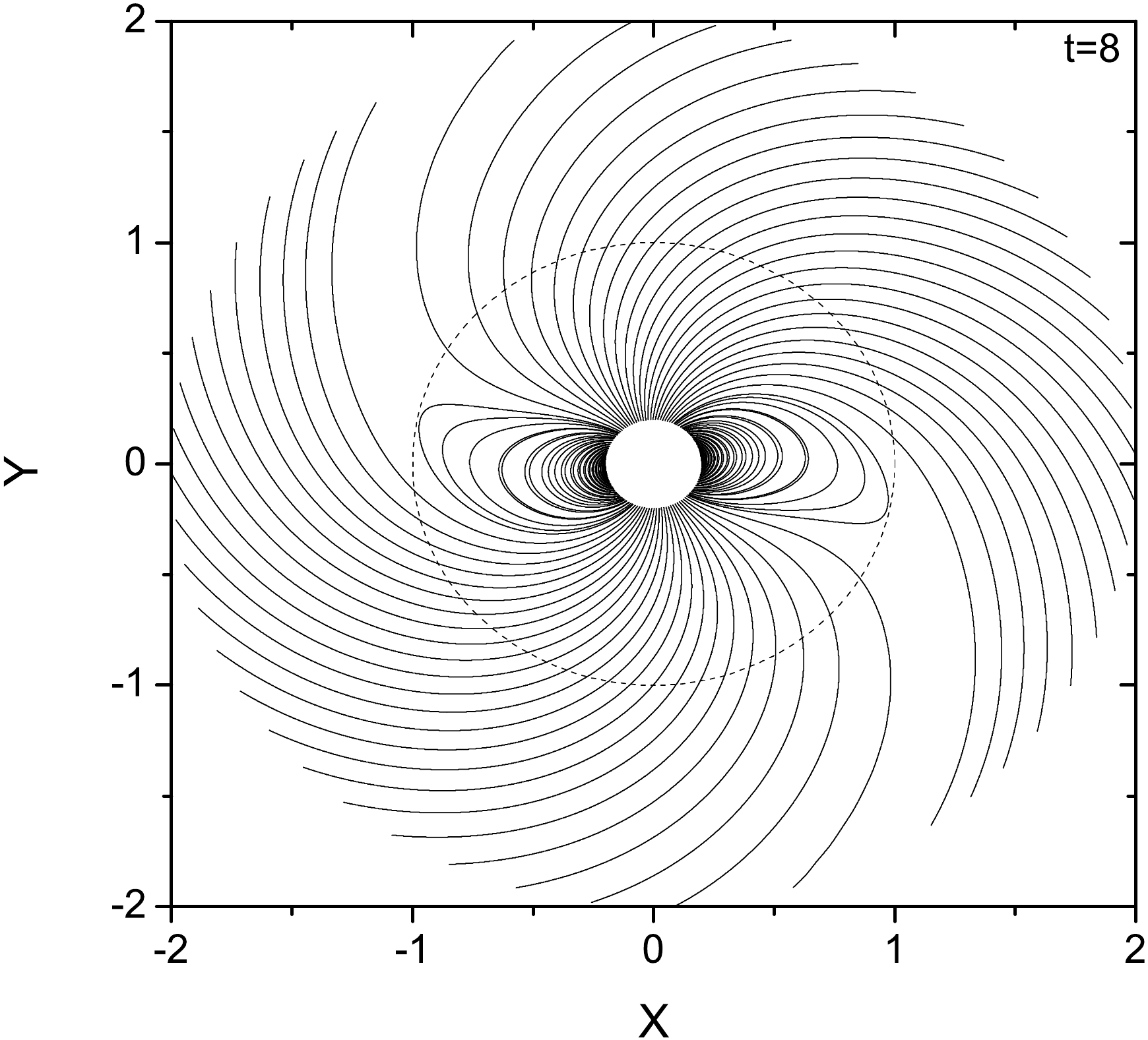}
\end{tabular}
\caption{Time sequence of equatorial field lines for an perpendicular force-free rotator. Light cylinder is shown as a black dashed circle. \label{fig1}}
\end{figure*}

We impose the exact inner boundary condition at the stellar surface with a rotating  electric field ${\bf {E}} = -( {\bf \Omega } \times \bf r ) \times \bf B$.
However, it is difficult to exactly handle with the outer boundary condition on a sphere of a finite radius. We use the characteristic compatibility method
described in \citet{can07} and \citet{pet12} to prevent the inward reflection from the outer boundary. No any spurious reflection from the outer boundary
is found in our simulation.
After the spectral transform from the real space to spectral space is performed, a set of partial differential equations are replaced by a larger set of ordinary differential equations with appropriate initial and boundary conditions. We use a third-order Adam-Bashforth scheme to advance the solution at each time step. In order to ensure the long-time stability of the algorithm and increase the convergent rate of the solution, it is necessary to filter the high-frequency mode to reduce the aliasing errors and the Gibbs phenomenon. The spectral filtering is performed at each time step. We use an eighth-order exponential filter in all directions by the expression
\begin{equation}
\sigma(\eta) = \textrm{e}^{-\alpha\,\eta^\beta}\;,
\end{equation}
where $\eta$ ranges from 0 to 1 and the parameter $\alpha\geq0$.
We implement the super spectral viscosity (SSV) filter in the radial direction explained in depth in \citet{boy98}. It preserves the boundary condition by adding  some
artificial viscosity.
For instance, in the radial direction, the filtering spectral coefficient is given as
\begin{equation}
B^r_{lm}(t,r)=\sum_{k=0}^{N_{r}-1}\sigma(\eta)B^r_{klm}(t) T_{k}(r)\;,
\end{equation}
where $\eta=k/(N_r-1)$ for $k\in[0,1,...,N_r-1]$.

\section{RESULT}

\subsection{Numerical setup}

In the following simulations, we adopt the following normalization: $B_{\rm \star}=\Omega=c=1$, where $B_{\rm \star}$ is the stellar surface magnetic field, $\Omega$ is the stellar angular velocity, and $c$ is the light velocity. Therefore, the light cylinder radial is $r_{\rm L}=c / \Omega=1$.

We extend the method developed by \citet{cao16} to 3D spherical coordinate system ($r$,\,$\theta$,\,$\phi$). The full set of Maxwell equations are solved by the pseudo-spectral method in spherical coordinates. The radial coordinate $r$ is expanded as a series of Chebyshev polynomials, and the angle coordinates $\theta$ and $\phi$ are expanded as VSHs. The computational domain is set to be $r\in (0.2 - 2)$ $r_{\rm L}$. In order to capture the current sheet, a SSV filter is used in the radial direction. A minimum resolution of $N_r \times N_{\theta} \times N_{\phi}=64 \times 32 \times 64$ is necessary to get a good accuracy.
We try a higher resolution with $N_r \times N_{\theta} \times N_{\phi}=128 \times 32 \times 64$, we find that the solution is similar and thus the radial dimensions can be resolved with the coarsest grid.

The magnetic field is initialized to be an rotating dipolar in vacuum \citep{mic99}
\begin{eqnarray}
B_r&=&2B_{\star}\frac{R_{\star}^3}{r^3}[\cos\alpha \cos\theta+\sin\alpha \sin\theta \cos(\phi-\Omega t)]\;,\\
B_{\theta}&=&B_{\star}\frac{R_{\star}^3}{r^3}[\cos\alpha\sin\theta-\sin\alpha\cos\theta\cos(\phi-\Omega t)]\;,\\
B_{\phi}&=&B_{\star}\frac{R_{\star}^3}{r^3}\sin\alpha\sin(\phi-\Omega t)\;,
\end{eqnarray}
where $\alpha$ is the magnetic inclination angle and is set to be $\alpha=\{0^\circ,15^\circ,30^\circ,45^\circ,60^\circ,75^\circ,90^\circ\}$.
Meanwhile, The electric field is set to be zero, ${\bf E}=0$, except for the stellar surface where we enforce the inner boundary condition with a corotating electric field. We let the system evolve for several rotational periods to reach a final steady solution.

\begin{figure}
\centerline{\epsfig{file=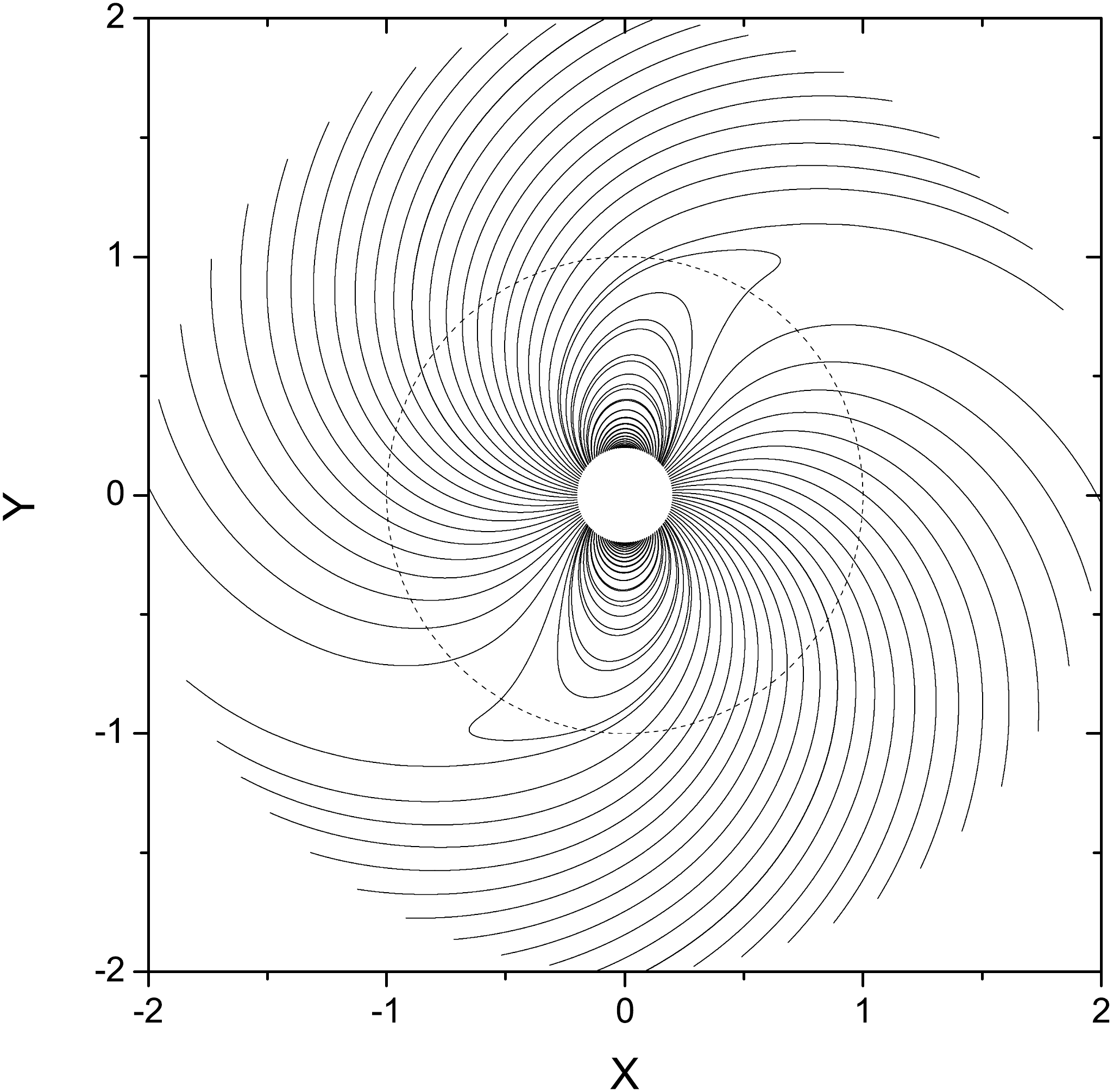, width=7.5cm}}
\caption{ Equatorial magnetic field lines for the perpendicular force-free rotator. \label{fig2} \protect \\ \protect \\ \protect \\}
\end{figure}

\begin{figure}
\centerline{\epsfig{file=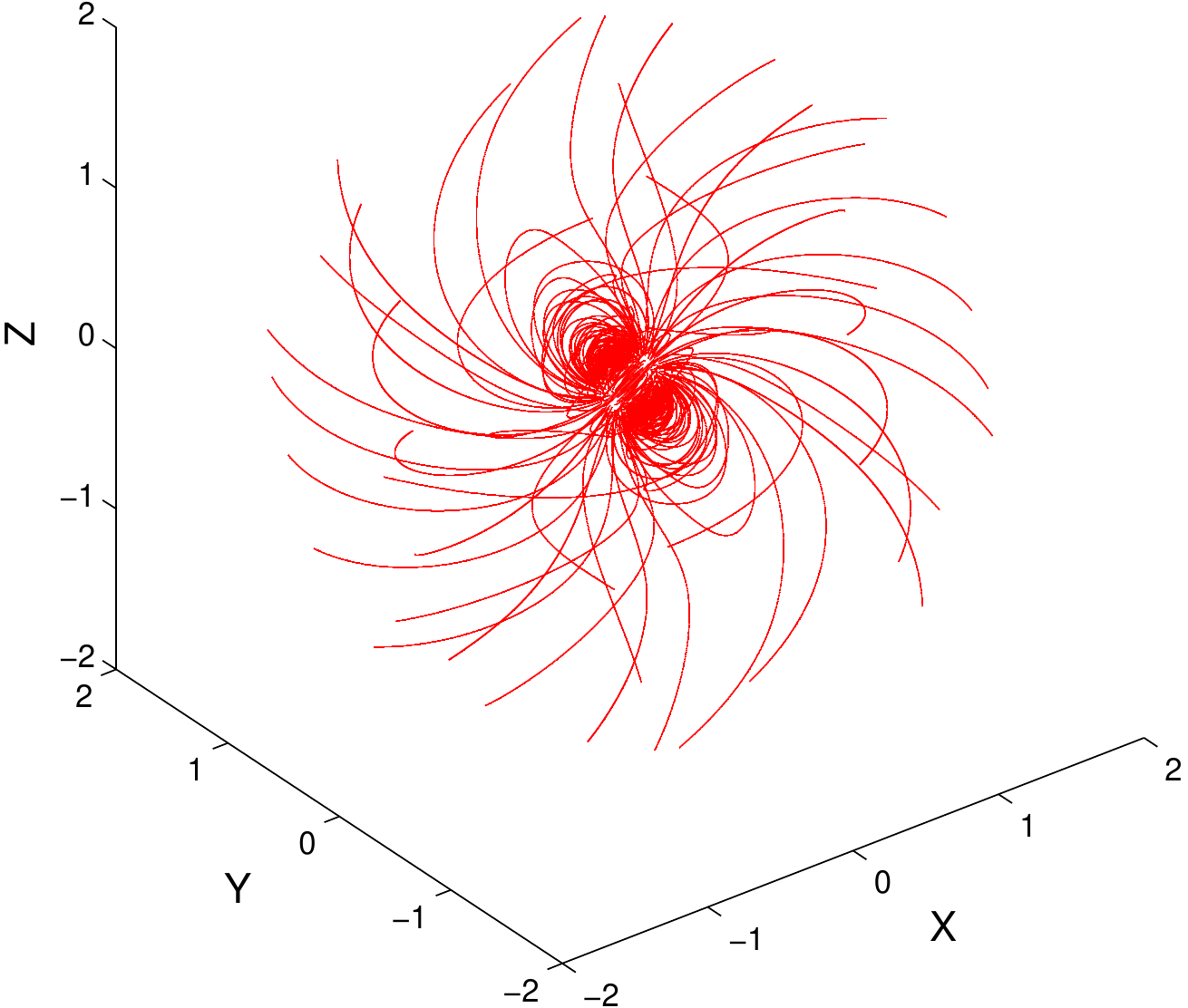, width=8cm}}
\caption{ 3D magnetic field lines for the inclined rotator with $\alpha=60^{\circ}$ \label{fig2}}
\end{figure}

\begin{figure}
\centerline{\epsfig{file=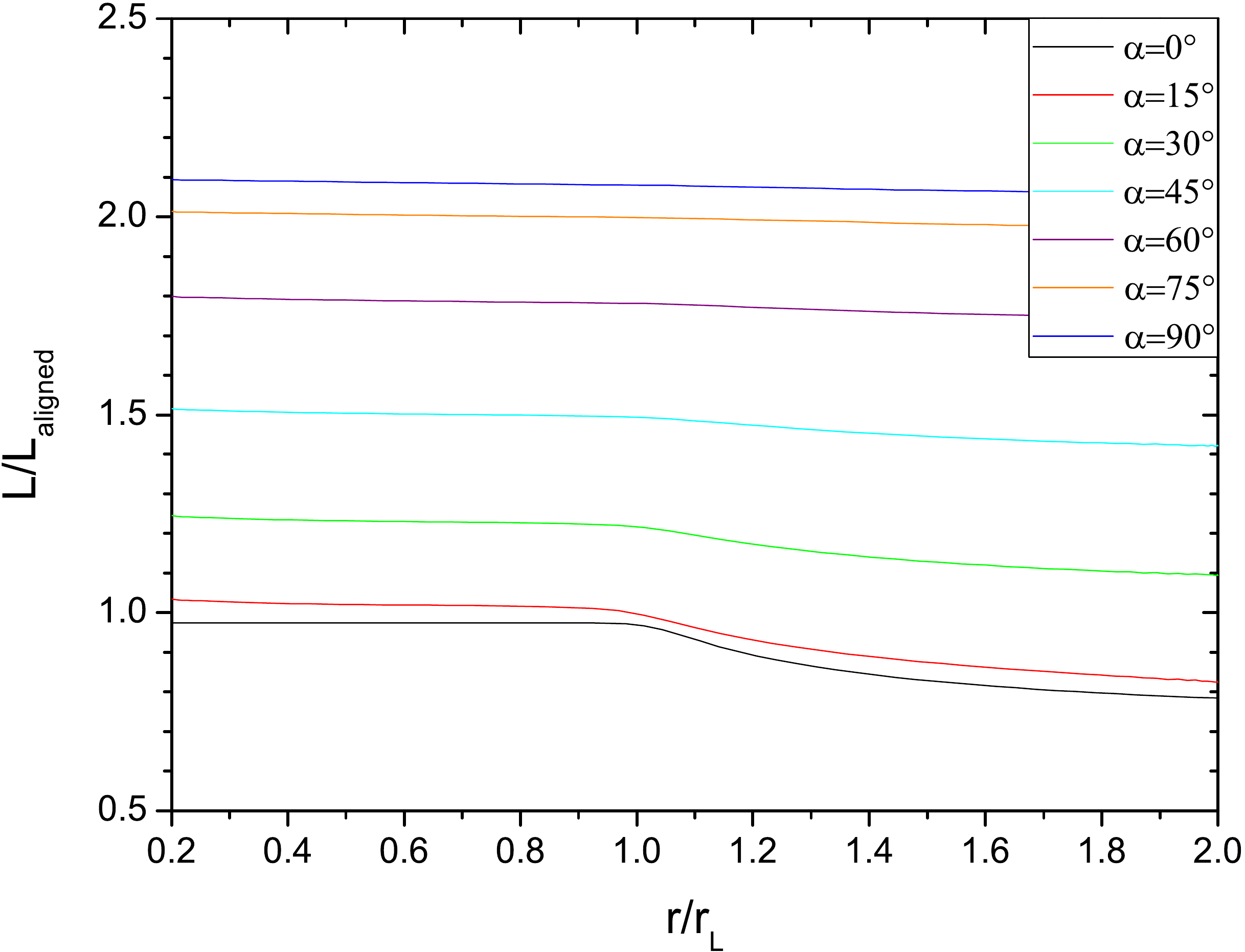, width=8cm}}
\caption{ The normalized Poynting flux $L/L_{\rm aligned}$ across the sphere of radial $r$ for different magnetic inclination angle $\alpha$. \label{fig4}}
\end{figure}

\begin{figure}
\centerline{\epsfig{file=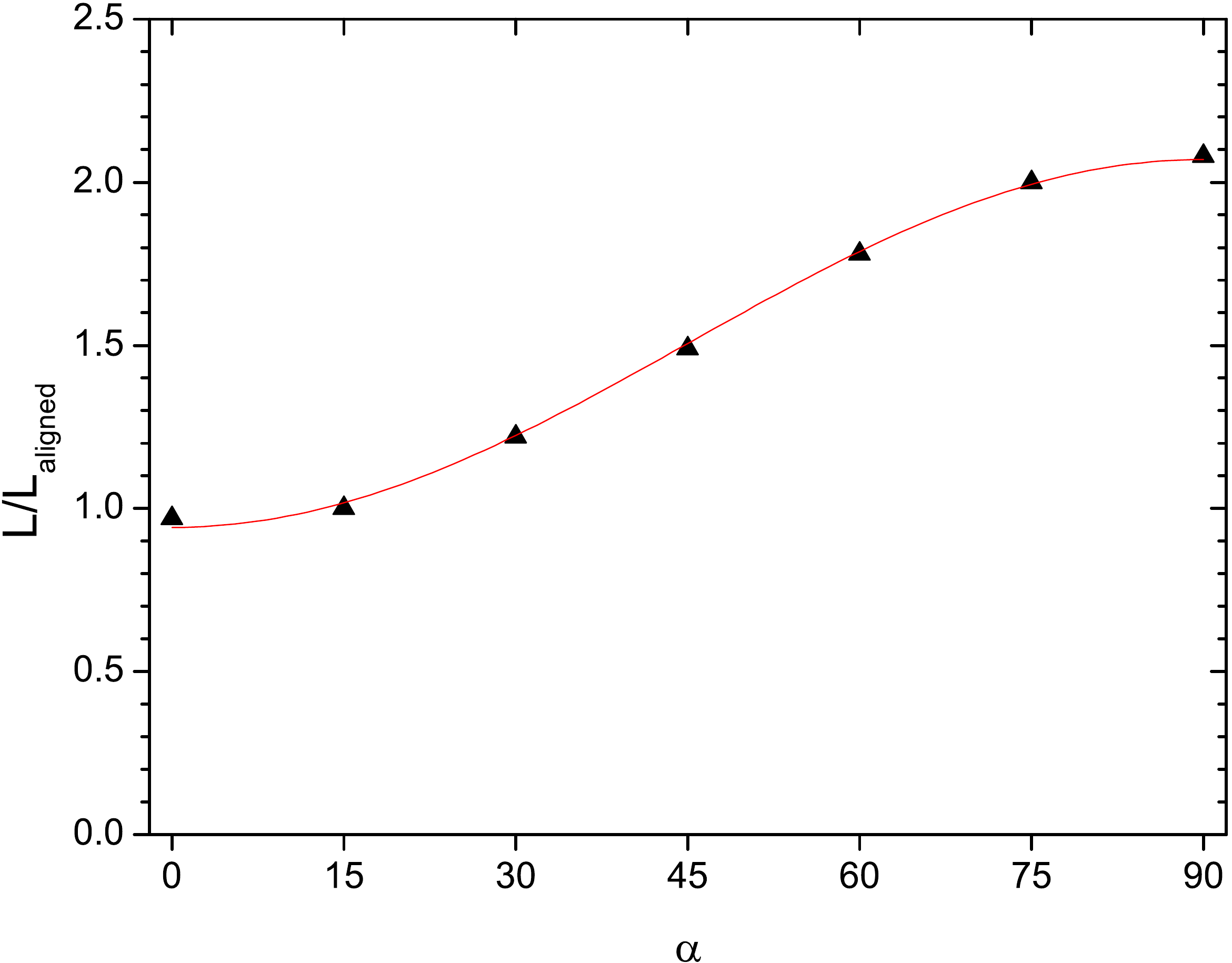, width=8cm}}
\caption{ The Poynting flux across the light cylinder $r_{\rm L}$ as a function of inclination angle $\alpha$. \label{fig5}}
\end{figure}

\subsection{Force-free pulsar magnetosphere ($\sigma\rightarrow\infty$)}

In our previous paper \citep{cao16}, we presented the structure of force-free pulsar magnetosphere for an aligned rotator. The solutions consist of regions of the closed and open field lines. However, some closed lines pass through the light cylinder because of artificial dissipation by spectral filtering. The Poynting flux inside the light cylinder remains constant and well agrees with the value obtained in previous works. However, dissipation beyond the light cylinder becomes significant due to the presence of the current sheet in the equatorial palne. Our solution is similar to that of \citet{par12} obtained by a spectral code.

Here, we extend  our pseudo-spectral code to 3D geometry by including the azimuthal fourier transform.
For the orthogonal rotator, visualization is easy because magnetic field lines are entirely contained in the equatorial plane.
In figure 1, we show a time sequence of the approach to steady state for the field lines in the equatorial plane. The approach towards steady state is similar to the aligned one, see \citet{cao16}. Our results are in qualitative agreements with  previous obtained results \citep{spi06,kal09,pet12}.
The final magnetic field lines in the equatorial plane are shown in figure 2. The field lines are reminiscent of the Deutsch field solution inside the light cylinder.
The field lines with spiral structure is seen in the regions of a strong magnetic field gradient outside the light cylinder.
Examples of 3D magnetic field line configuration for an oblique rotator with $\alpha=60^\circ$ is shown in the figure 3. The magnetic topology is reminiscent of the CKF solution with closed and open regions. The field lines are similar to the inclined split-monopole solution far away the star. Compared to the aligned rotator, The current sheet outside the light cylinder is more resolved by our current grid resolution.

The Poynting flux integrated over a sphere of a radial distance $r$ is given by
\begin{equation}
L=\int r^2 d\Omega_{\rm s} \,{\bf S \cdot}{\bf e}_{r}\;,
\end{equation}
where $ {\bf {S}}= {\bf {E \times B}}$ is the Poynting vector, $d \Omega_{\rm s}$ is the infinitesimal solid angle, and $\Omega_{\rm s}$ is the full sky angle of $4\pi$ sr.
The Poynting flux for the force-free aligned rotator is given by \citep{spi06}
\begin{equation}
L_{\rm aligned}=\frac{B_{\star}^2 R_{\star}^6 \Omega_{\star}^4}{c^3}\;.
\end{equation}
The normalized Poynting flux $L/L_{\rm aligned}$ across the sphere of radial $r$ is shown in figure 4.
Inside the light cylinder, the Poynting flux is nearly constant with radial for all inclination angle $\alpha$, which is expected by the conservation of energy. A good accuracy can be obtained with a reasonable number of grid points $N_{\theta}=32$. Outside the light cylinder, the Poynting flux decreases with radius. This is because of the presence of the current sheets where discontinuities in the magnetic field appear. The current sheets are a general features of force-free electrodynamics. The discontinuities in the current sheets are smeared by  our filtering procedure. When inclination angle $0^{\circ}\leq\alpha<60^{\circ}$, dissipation of the Poynting flux become significant outside the light cylinder, and the energy dissipation decreases with increasing inclination angle $\alpha$. When inclination angle $\alpha\geq60^{\circ}$, the energy dissipation become negligible. The Poynting flux remains constant and  energy is conserved to a good accuracy in the whole computational regions.
\begin{figure*}
\begin{tabular}{cccccc}
  \includegraphics[width=5.5cm,height=5.5cm]{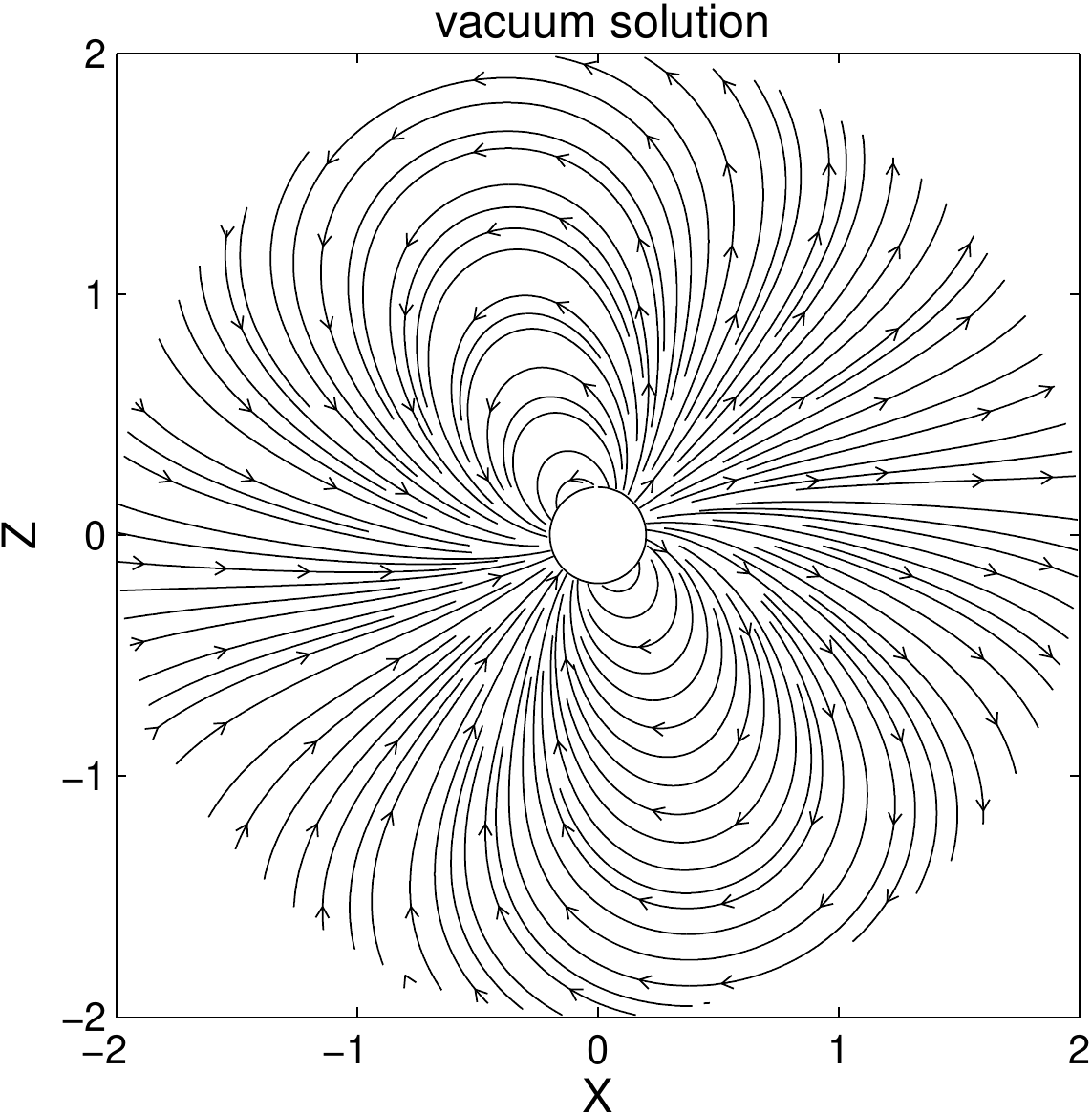} \qquad
  \includegraphics[width=5.5cm,height=5.5cm]{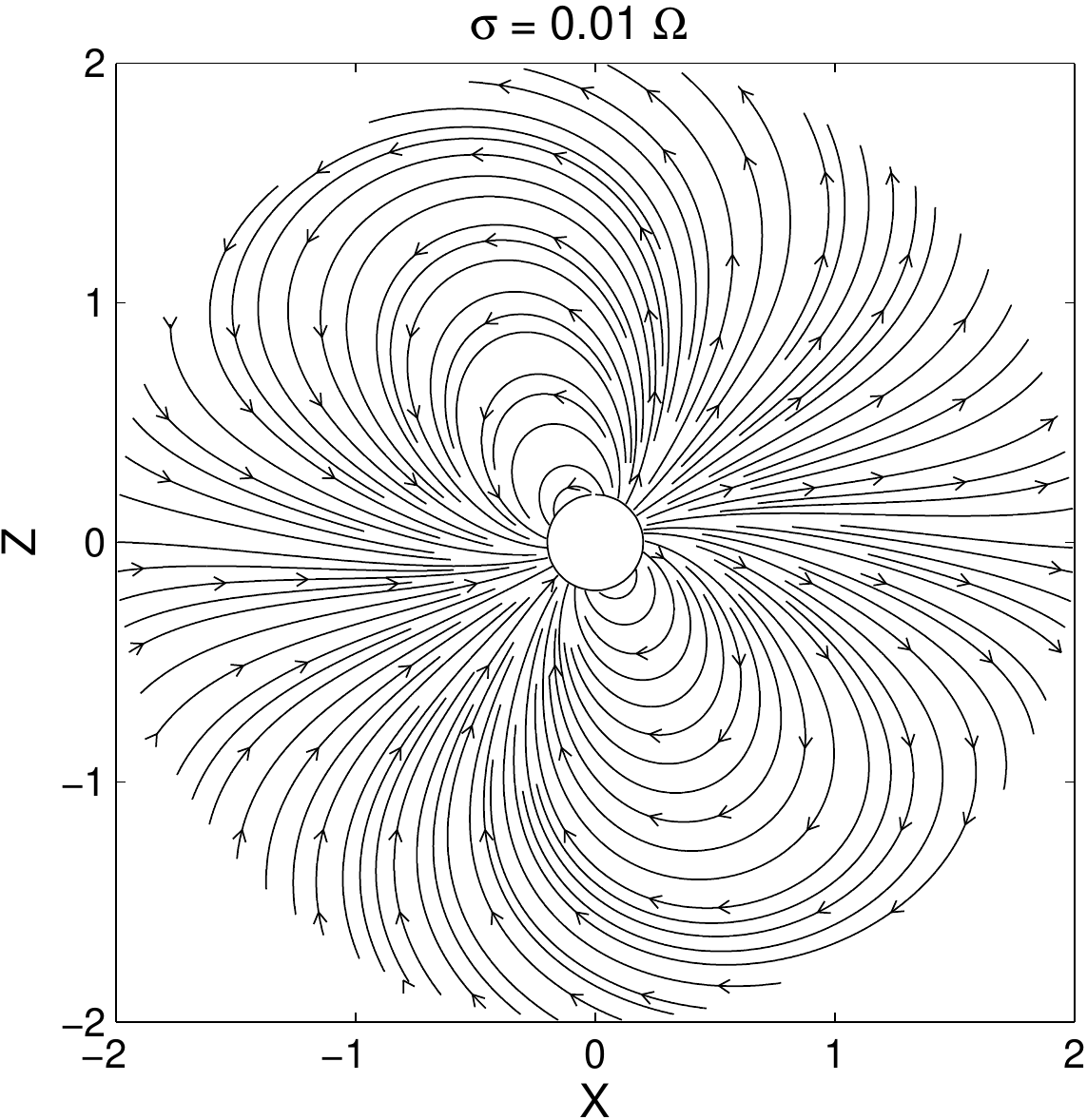} \qquad
  \includegraphics[width=5.5cm,height=5.5cm]{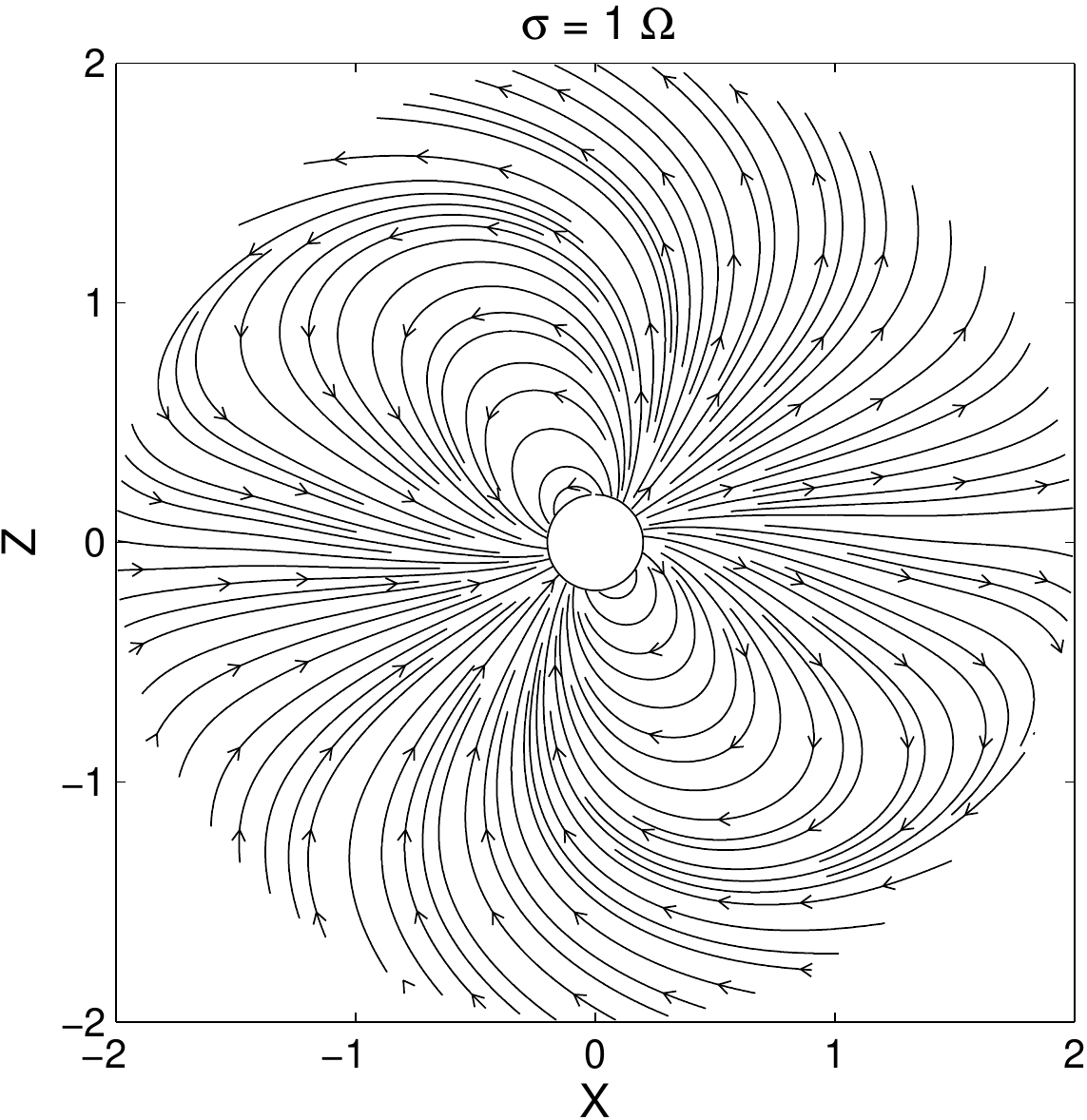} \\
  \includegraphics[width=5.5cm,height=5.5cm]{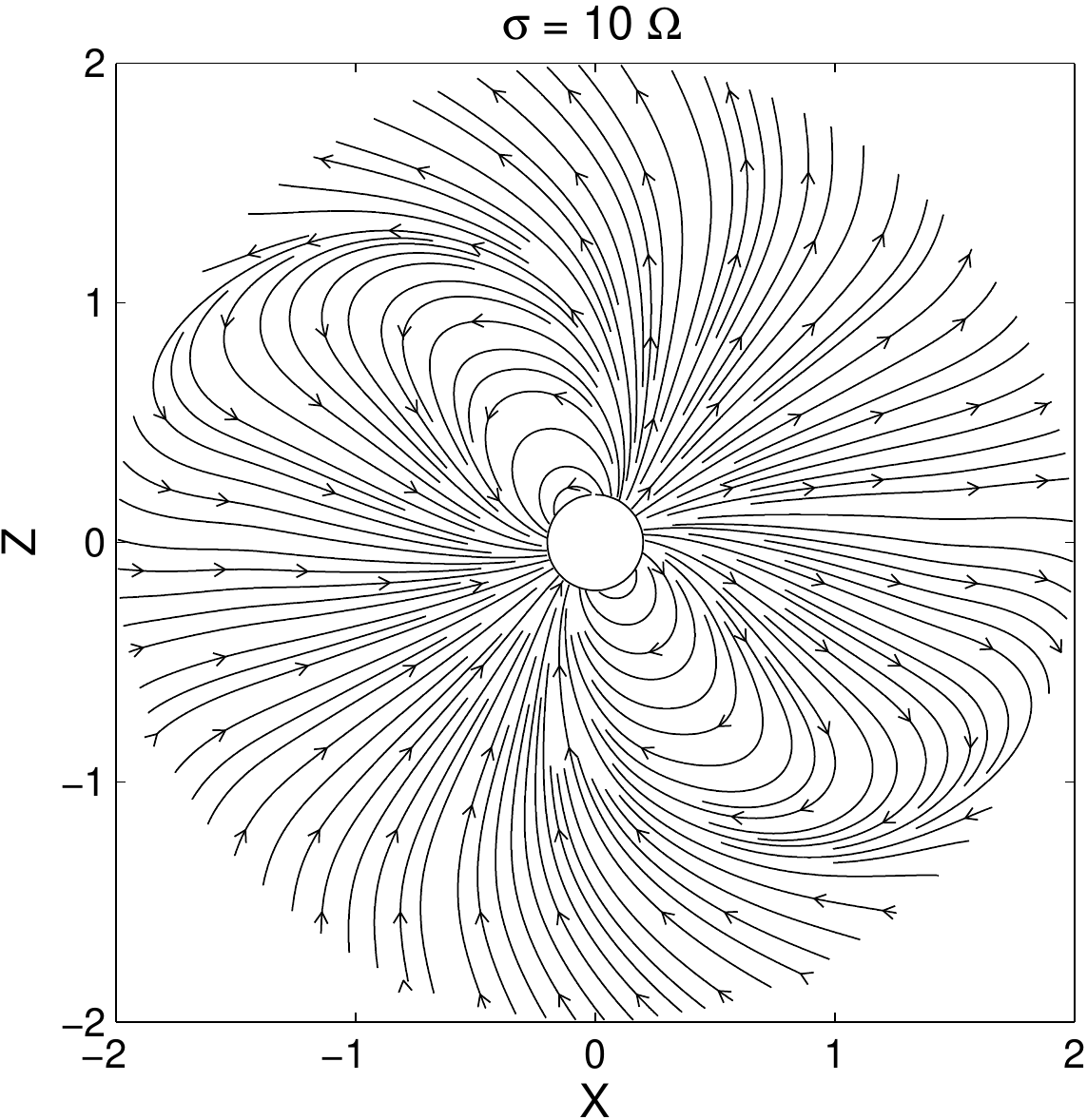} \qquad
  \includegraphics[width=5.5cm,height=5.5cm]{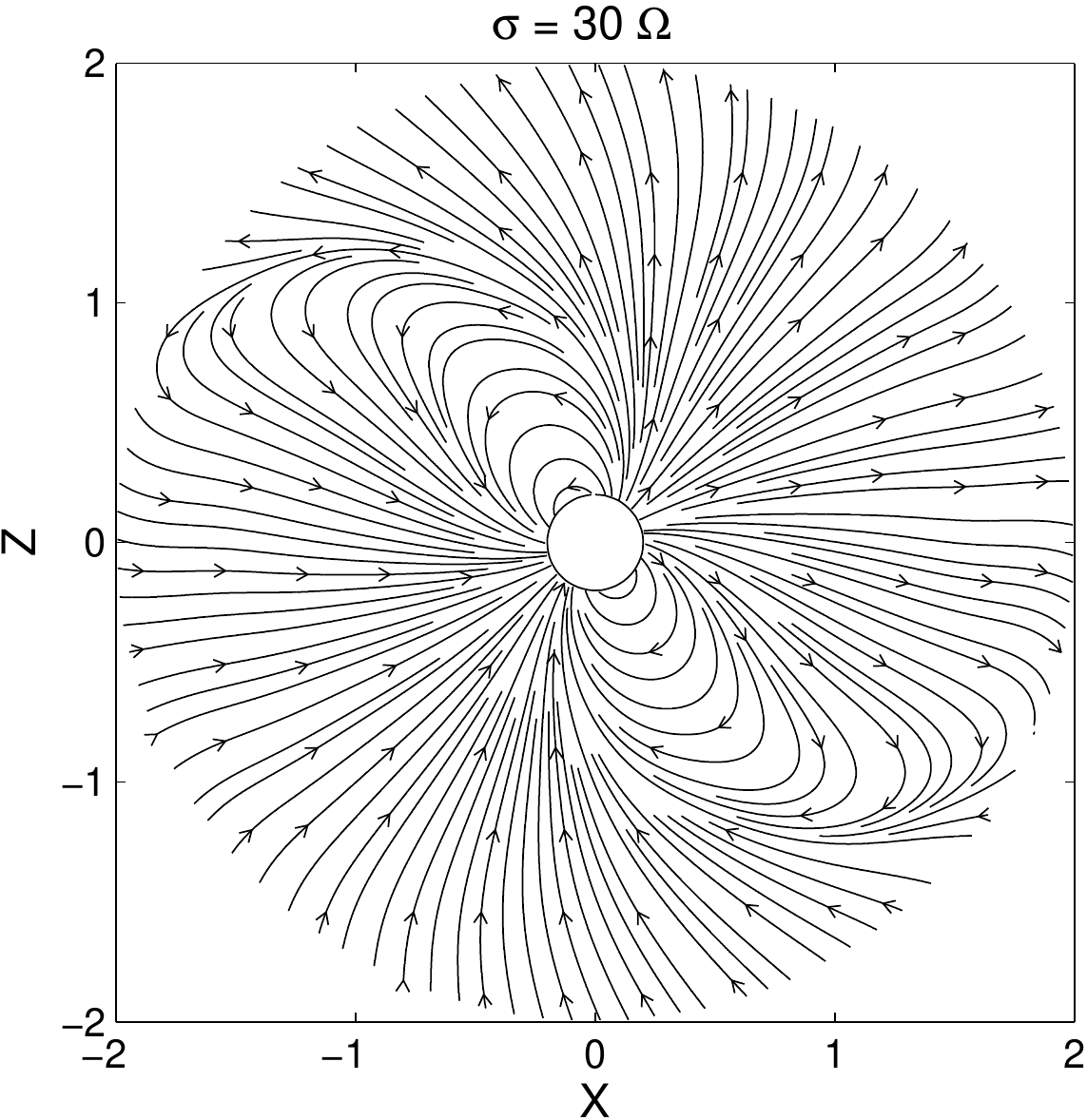} \qquad
  \includegraphics[width=5.5cm,height=5.5cm]{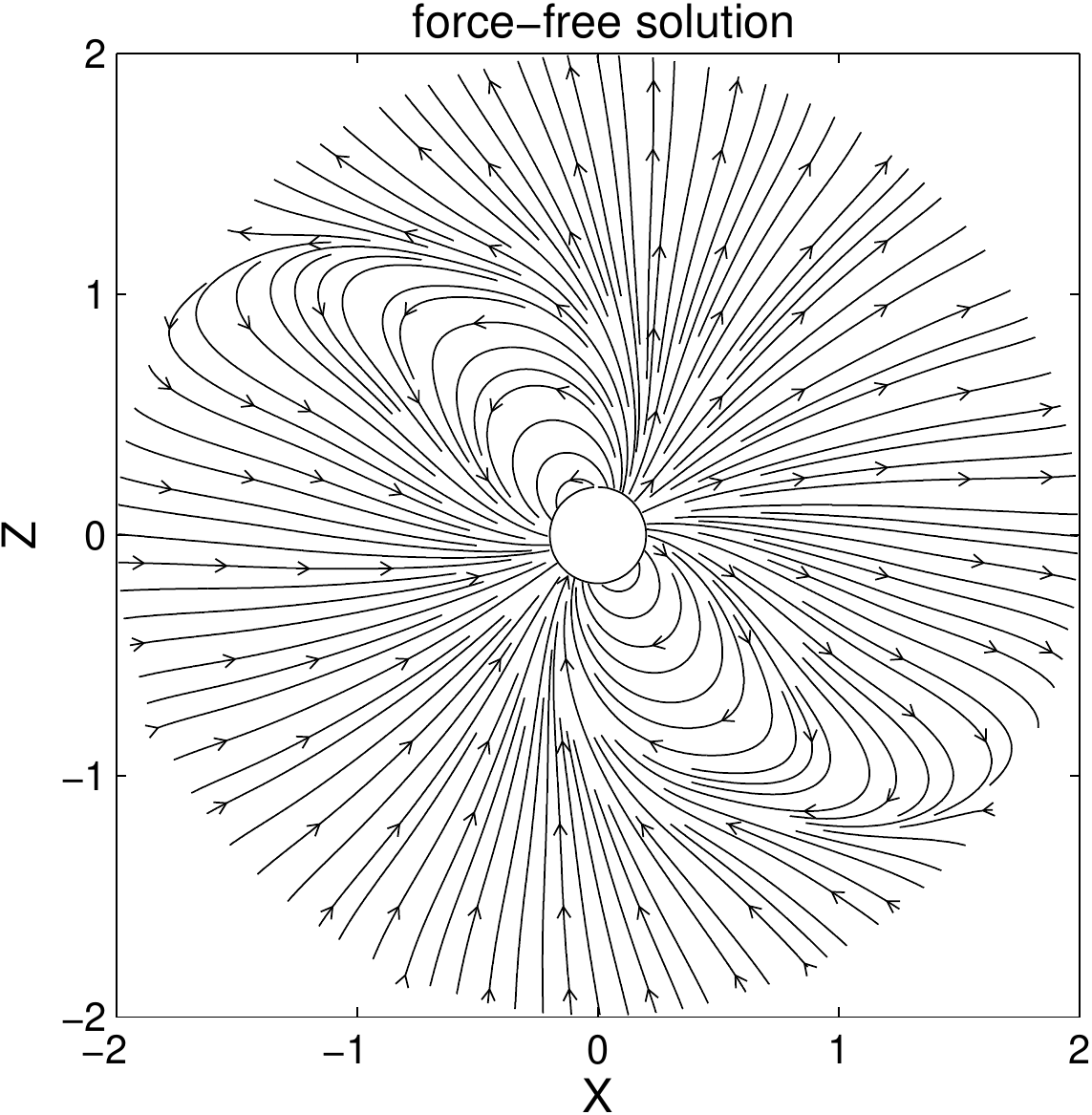} \\
\end{tabular}
\caption{Magnetic field lines in the x-z plane for a $60^{\circ}$ inclined rotator with different conductivity $\sigma$.}
\end{figure*}
Displacement currents are not present in the equatorial current sheet for the aligned rotator, but the displacement currents increase with increasing inclination
angle $\alpha$ and dominates over the conduction current when inclination angle $\alpha\geq60^{\circ}$. It seems that the current sheet can be better resolved by our coarse grid due to the presence of a strong displacement currents. The Poynting flux at the light cylinder as a function of inclination angle $\alpha$ is fitted by
the expression
\begin{equation}
\frac{L}{L_{\rm aligned}}=k_1+k_2\sin^2{\alpha}\;,
\end{equation}
where $k_1$ and $k_2$ are constants. The fit results are shown in figure 5. The best fit parameters are $k_1=0.94$ and $k2=1.12$.
Our result is in qualitative agreement with that of \citet{spi06}, although the full expressions for the current density are taken into account in our simulations.

\subsection{Resistive pulsar magnetosphere (finite $\sigma$)}

In the force-free regime, the accelerating electric field is shorted out by  abundant plasma. They do not accommodate the production of radiation, in disagreement with the observations. The plasma resistivity can allow the accelerating electric field. Here, we extend the force-free solution to the resistive solution with a finite conductivity $\sigma$. We perform a series of simulation for different conductivities $\sigma=\{0.01,\, 0.1,\, 0.3,\, 1,\, 3,\, 5,\, 10,\, 15,\, 30,\, 60\} \,\, \Omega$.
Magnetic field lines in the x-z plane for a $60^{\circ}$ inclined rotator with different conductivities are depicted in figure 6.
When $\sigma=0.01 \, \Omega$, the global structure of the field lines is very similar to that of the vacuum solution. As the conductivity $\sigma$ increases, the resistive solution gradually deviates from the Deutsch vacuum solution due to magnetospheric current. A fraction of open field lines increase with increasing conductivity  and the solution approaches the force-free solution for high value of $\sigma$. When $\sigma=30 \, \Omega$, the field line configuration
is very similar to that of the force-free solution.

The Poynting flux as a function of magnetic inclination angle $\alpha$ is given by
\begin{equation}
\frac{L}{L_{\rm aligned}}=\left\{\begin{array}{ll}
\frac{2}{3}\sin^2 \alpha   \quad                               &\mbox{the vacuum solution}\;,\\
1+\sin^2 \alpha            \quad                               &\mbox{the force-free solution}\;. \\
\end{array}
\right.
\end{equation}
It is expected that the resistive solutions should lie between the vacuum and force-free limits.
We check the dependence of the Poynting flux on magnetic inclination angle $\alpha$ for $\sigma=\{0.01,\, 0.1,\, 0.3,\, 1,\, 3,\, 5,\, 10,\, 15,\, 30,\, 60\} \,\, \Omega$. For comparison, the Deutsch vacuum and force-free solution are shown as the dashed lines. The results are shown in figure 7. When $\sigma=0.01\,\Omega$, the Poynting flux is very close to that of the vacuum field. The Poynting flux increases with the increasing conductivity and approaches the force-free solution. When $\sigma=60\,\Omega$, the Poynting flux is close to that of the force-free field, especially for the high obliquity rotator ($\alpha\geq60^{\circ}$). A clear transition is seen from the vacuum limit to the force-free limit with the increasing conductivity. Our results are in qualitative agreement with the previous obtained results \citep{li12a,kal12}.

The dissipation power is defined by the expression \citep{kal12}
\begin{equation}
\dot{E}_{d}=\int_{r_1}^{r_2}{\bf J}\cdot{\bf E} \, dV = \int_{r_1}^{r_2} \sigma\,E^2_{\parallel} \, dV .
\end{equation}
The spectral filtering in the radial direction slightly changes the inner boundary condition. Therefore, we set $r_{1}=0.4 \, r_{\rm L}$ and $r_{2}=2 \, r_{\rm L}$ in order to more precisely estimate the dissipation power. The dissipation power $\dot{E}_{d}$, normalized to the Poynting flux $L_{1}$ measured at radial $r_{1}$, as a function of $\sigma$ is shown in figure 8. The dissipation power goes to $0$ as $\sigma$ goes to $0$ or $\infty$. This is because the vacuum and force-free solution is dissipationless (see equation 21). The dissipation power is maximum at intermediate conductivities of $\sigma/\Omega\sim1$. The maximum dissipation occurs at higher values of $\sigma$ as the inclination angle $\alpha$ decreases.
Our results are similar to that of \citet{kal12}. The parallel electric field  for a $60^{\circ}$ inclined  rotator with $\sigma=15\,\Omega$ is shown in figure 9. We see that most of dissipation takes place in the polar region inside the light cylinder and in the current sheet outside the light cylinder for the solutions of high conductivity.

\begin{figure}
\centerline{\epsfig{file=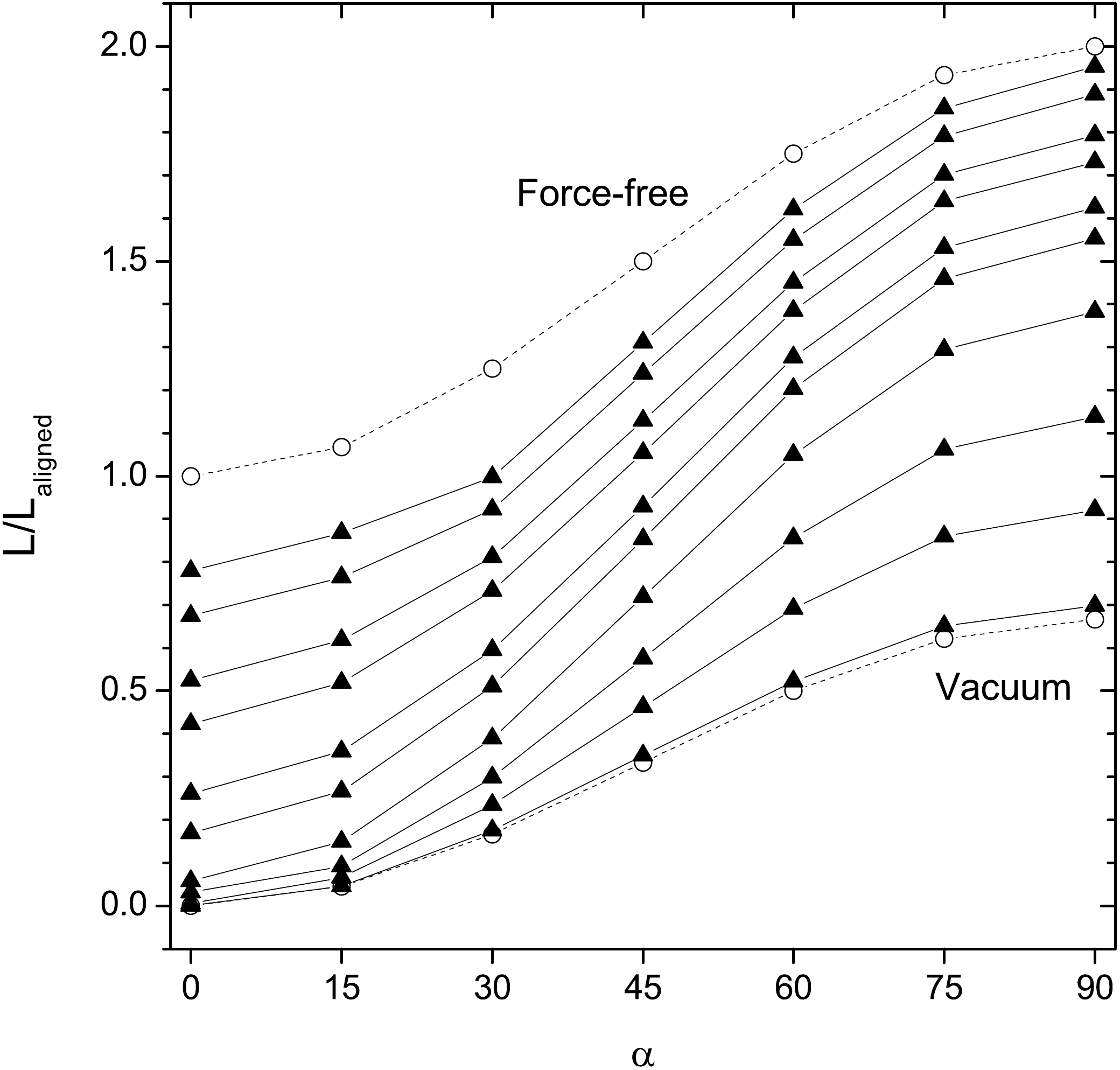, width=8cm}}
\caption{ The Poynting flux dependence on inclination angle $\alpha$ for the vacuum solution, a sequence of resistive solution with $\sigma$ from 0.01 $\Omega$ to 60 $\Omega$ and the force-free solution (from bottom to top).}
\end{figure}

\begin{figure}
\centerline{\epsfig{file=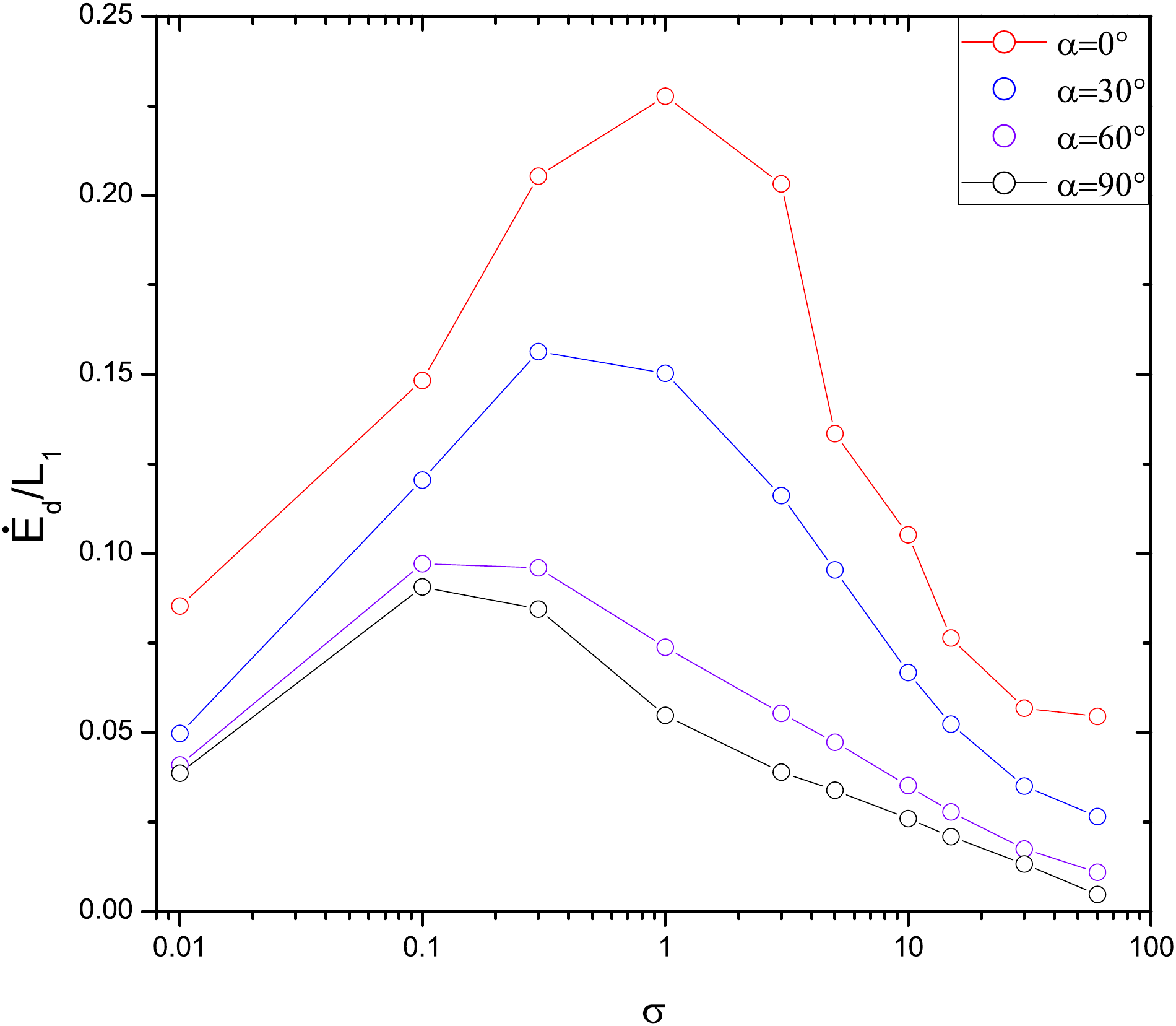, width=8cm}}
\caption{The normalized dissipation power $\dot{E}_{d}/L_1$ as a function of $\sigma$ for different inclination angle $\alpha$.}
\end{figure}

\begin{figure}
\centerline{\epsfig{file=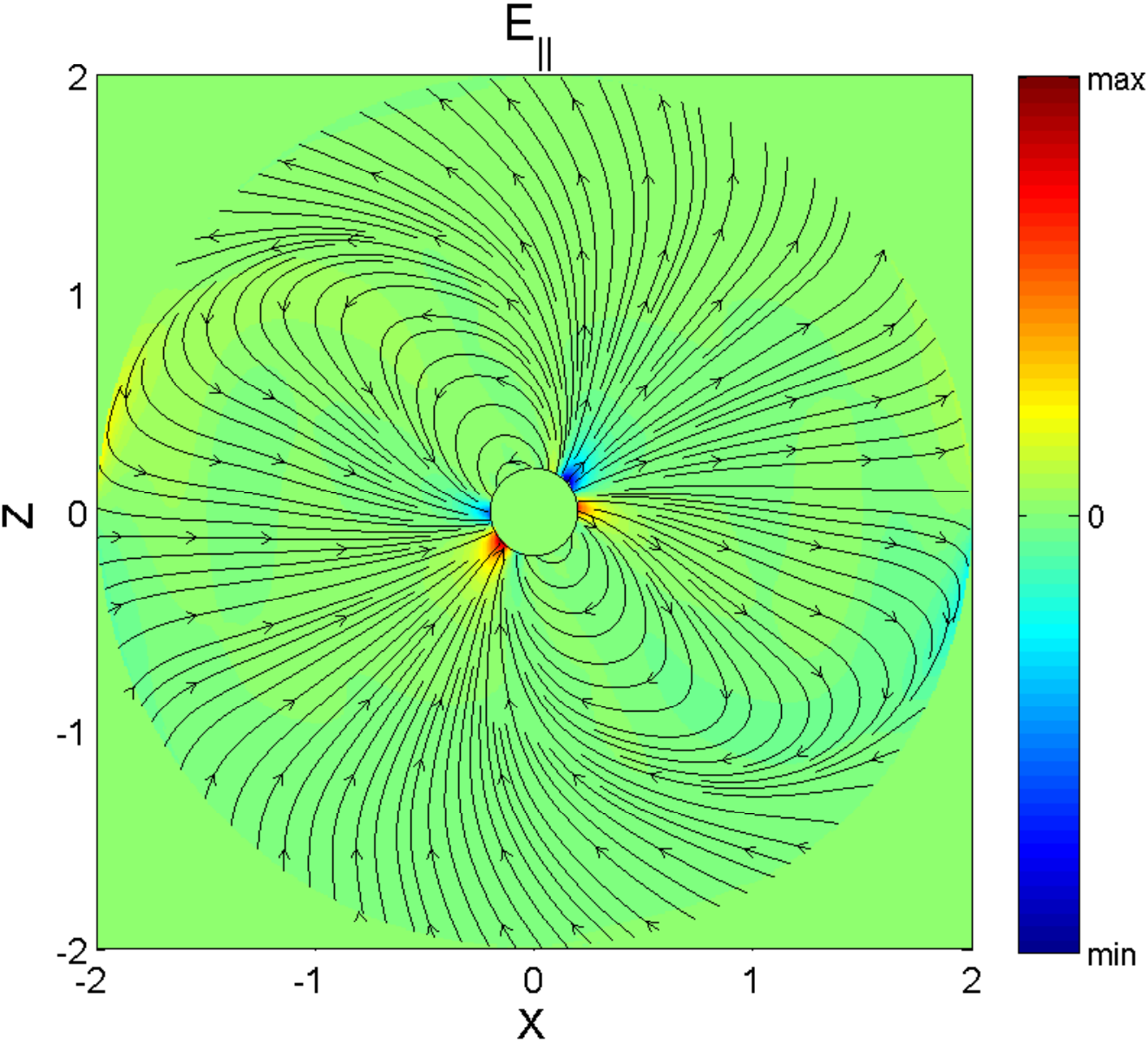, width=8cm}}
\caption{The parallel electric field $E_{\parallel}$ for a $60^{\circ}$ inclined  rotator with $\sigma=15\,\Omega$.}
\end{figure}

\section{DISCUSSION AND CONCLUSION}

In this paper, we have solved a set of the time-dependent Maxwell equations by the pseudo-spectral code and investigated the pulsar magnetosphere with a finite resistivity. Through extending the study of \citet{cao16} to 3D oblique rotator, at first we have simulated the structures of oblique pulsar magnetosphere in the force-free approximation and retrieved the force-free spin-down luminosity, which is in qualitative agreement with that given by \citet{spi06}. Finally we have simulated the structures of oblique pulsar magnetosphere in the vacuum, finite $\sigma$, and the force-free cases, and produced a set of solutions that smoothly transition from the Deutsch vacuum solution to the force-free solution with increasing conductivity, which are in qualitative agreement with the previous results given by \citet{li12a} and \citet{kal12}.

Inside the light cylinder, A good accuracy can be obtained by our current grid resolution. However, the dissipation of energy becomes significant outside the light cylinder due to the presence of discontinuities in the current sheets, especially for low obliquity rotator ($\alpha<60^{\circ}$). The only way to reduce the dissipation is to increase the grid resolution. However, the current available computational resources cannot allow us to perform the very high resolution simulations. The simulations presented here were obtained on a single processor. We plan to improve the code by writing a Message Passing Interface (MPI) version in the near future. Then, we will use the resistive solution to model the Fermi observed $\gamma$-ray spectra and light curves with higher resolution simulation.
Particle inertia and pressure are missing in the force-free regime, which is crucial for understanding the realistic magneto-spherical structures and  modeling the realistic $\gamma$-ray light curves.
Therefore, we will relax the ideal MHD approximation by including particle inertia and pressure.
It is straightforward to extend our code to solve a full set of MHD equations.

\section*{Acknowledgments}
We thank the anonymous referee for valuable comments and suggestions.
We would like to thank J\'{e}r$\hat{\rm o}$me P\'{e}tri, Kyle Parfrey
and Cong Yu for some useful discussions. This work is partially supported by the National Natural Science Foundation of China (NSFC 11433004, 11103016, 11173020), the Doctoral Fund of the Ministry of Education of China (RFDP 20115301110005), and the Top Talents Programme of Yunnan Province (2015HA030).


\bibliography{refernces}


\end{document}